\documentclass[aps,prl,twocolumn,superscriptaddress,nobibnotes]{revtex4-2}

\usepackage[british]{babel}
\usepackage[babel=true]{csquotes}
\usepackage{graphicx}
\usepackage{float}

\usepackage[colorlinks=true,linkcolor=blue,citecolor=blue,urlcolor=blue]{hyperref}

\usepackage{amsmath}
\usepackage{amssymb}
\usepackage{mathtools}
\usepackage{booktabs}
\usepackage{dsfont}

\usepackage[nopostdot,style=super,nonumberlist,toc,nogroupskip]{glossaries}
\glsdisablehyper

\usepackage{siunitx}
\DeclareSIUnit\sq{\ensuremath{\Box}}
\usepackage[colorinlistoftodos,prependcaption,textsize=tiny]{todonotes}

\usepackage{color,soul}

\graphicspath{{../images/}}

\newacronym{soi}{SOI}{silicon-on-insulator}
\newacronym{hom}{HOM}{Hong-Ou-Mandel}
\newacronym{snspd}{SNSPD}{superconducting nanowire single-photon detector}
\newacronym{spdc}{SPDC}{spontaneous parametric down-conversion}
\newacronym{ppktp}{ppKTP}{periodically poled potassium titanyl phosphate}
\newacronym{hwp}{HWP}{half-wave plate}
\newacronym{qwp}{QWP}{quarter-wave plate}
\newacronym{pbs}{PBS}{polarising beam splitter}
\newacronym{mmi}{MMI}{multi-mode interferometer}
\newacronym{gc}{GC}{grating coupler}
\newacronym{mzi}{MZI}{Mach-Zehnder interferometer}
\newacronym{2dc}{2DGC}{two-dimensional grating coupler}
\newacronym{fst}{FST}{full quantum state tomography}
\newacronym{mle}{MLE}{maximum likelihood estimation}
\newacronym{box}{BOX}{buried oxide}
\newacronym{tox}{TOX}{top oxide}
\newacronym{ghz}{GHZ}{Greenberger-Horne-Zeilinger}
\newacronym{mbqc}{MBQC}{measurement-based quantum computing}
\newacronym{pic}{PIC}{photonic integrated circuit}
\newacronym{te}{TE}{transverse electric}
\newacronym{ce}{CE}{coupling efficiency}
\newacronym{ps}{PS}{phase shifter}

\newcommand{\affiliationBARZ}{\affiliation{Institute for Quantum Information and Technology,
 University of Stuttgart, 70569 Stuttgart, Germany.}}
\newcommand{\affiliationIQST}{\affiliation{Center for Integrated Quantum Science and Technology (IQST),
 University of Stuttgart, 70569 Stuttgart, Germany.}}

\newcommand{\state}[1]{|{#1}\rangle}
\makeatletter
\newsavebox{\@brx}
\newcommand{\llangle}[1][]{\savebox{\@brx}{\(\m@th{#1\langle}\)}%
  \mathopen{\copy\@brx\kern-0.5\wd\@brx\usebox{\@brx}}}
\newcommand{\rrangle}[1][]{\savebox{\@brx}{\(\m@th{#1\rangle}\)}%
  \mathclose{\copy\@brx\kern-0.5\wd\@brx\usebox{\@brx}}}
\makeatother

\newcommand{\exval}[1]{\langle #1 \rangle}
\newcommand{\inequal}[0]{\mathcal{B}}
\newcommand{\p}{\mathord{+}} 
\newcommand{\m}{\mathord{-}} 

\begin{document}

\title{Measurement-Based Quantum Computing on a Photonic Chip}

\author{Jeldrik Huster}
    \affiliationBARZ{}
    \affiliationIQST{}
\author{Louis L. Hohmann}
    \affiliationBARZ{}
    \affiliationIQST{}
\author{Kevin Edelmann}
    \affiliation{Institut für Mikroelektronik Stuttgart (IMS CHIPS), 70569 Stuttgart, Germany.}
\author{Stefanie Barz}
    \email[Corresponding author. Email: ]{barz@fmq.uni-stuttgart.de}
    \affiliationBARZ{}
    \affiliationIQST{}

\date{\today}
\begin{abstract}
    Integrated photonics provides a scalable platform for quantum information processing. 
    In this context, \gls{mbqc} offers an attractive approach in which quantum computation is realised by adaptive measurements on highly entangled graph states, circumventing the need for deterministic photon-photon interactions. 
    Here, we demonstrate \gls{mbqc} on an integrated silicon photonic chip capable of generating photonic graph states with up to four qubits. 
    We achieve fidelities of $F_\text{Star}=\qty{83.5\pm 1.8}{\percent}$ and $F_\text{Lin}=\qty{75.6 \pm 1.1}{\percent}$ for four-photon star and linear graph states, respectively. 
    We use these resource states to implement \gls{mbqc}-based single- and two-qubit gates and to demonstrate Grover's search algorithm and the Deutsch–Jozsa algorithm. 
    These results establish the feasibility of reconfigurable four-photon \gls{mbqc} on an integrated photonic platform and provide a foundation for future larger-scale implementations.
\end{abstract}

\maketitle

\section{Introduction}
    Photonic systems are excellent quantum information carriers, owing to their inherently low decoherence, high-speed propagation, and capacity for seamless connectivity. 
    Quantum information can be encoded into a multitude of photonic degrees of freedom, such as polarisation, time bins, and path, thereby providing a versatile framework for quantum information processing \cite{couteau2023}.
    Universal quantum computation requires a set of single-qubit gates in combination with two-qubit gates, such as the controlled-NOT gate or controlled-$Z$ gate. Whilst single-qubit gates are easy to implement in photonic systems, two-qubit gates are more challenging as photons do not directly interact.     
    \Acrfull{mbqc} circumvents this limitation by shifting the challenge from gate operations to a sequence of adaptive measurements performed on a highly entangled, pre-prepared cluster state~\cite{raussendorf2001, briegel2009}.
    A recent evolution of this framework is fusion-based quantum computing (FBQC), where small resource states are fused together via probabilistic \textit{fusion} measurements to realise fault-tolerant quantum computation~\cite{Browne2005, PsiQFusion2023}. 
    Consequently, \gls{mbqc} shifts the challenge of photonic quantum computing to the efficient generation of entangled resource states and the implementation of adaptive measurements.
    The \gls{mbqc} framework has been demonstrated in various proof-of-principle experiments in bulk optical systems. These experiments have successfully shown single- and two-qubit operations, quantum algorithms~\cite{walther2005, prevedel2007, tame2007, lee2012, Barz2012} with the additional implementation of active, high-speed feed-forward logic~\cite{prevedel2007, vallone2008}.\\
    The transition to advanced quantum information processing requires a significant increase in qubit numbers 
    and thus miniaturisation of the underlying physical footprint to support compact, robust implementations of larger, more complex circuit architectures.
    Such scaling can be realised with integrated optics and photonic integrated circuits replacing the bulk optical setups~\cite{zhu2026}. \\
    Various material platforms are actively developed to reach the requirements for large-scale photonic quantum information processing. 
    Advanced platforms like \gls{soi} and silicon nitride ($\text{Si}_3\text{N}_4$) offer mature fabrication  with low propagation losses~\cite{PsiQComponents2025}. 
    More recently, thin-film lithium niobate ($\text{LiNbO}_3$) and lithium tantalate (LiTaO$_3$) have gained significant interest due to their strong electro-optic effect and hence the potential for the fast switching speeds necessary for real-time adaptive feed-forward measurements~\cite{sund2023, zhang2026, wang2024b, cai2026}.
    Integrated photonics enables the generation, manipulation, routing and measurement of single photons on a single chip structure, with further scaling via efficient optical fibre interconnections~\cite{PsiQComponents2025}.
    Photonic integrated circuits (PICs) have been used for quantum information processing, including the postselected
    and the heralded generation of resource states, gate-based computations, and interconnections between chips ~\cite{Adcock2019a, llewellyn2020, vigliar2021, Bao2023, Lee2024, Pont2024, huang2024, chen2024a, chen2024, maring2024,
    forbes2025, carolan2015, qiang2018, chi2022}. 
    Experiments in integrated \gls{mbqc} have demonstrated single-qubit gate operations and the Grover's search algorithm utilising two-photon hyperentanglement~\cite{vigliar2021, zhang2023, ciampini2016}.  
    Furthermore, the generation of integrated graph states on continuous-variable encoded qubits has been demonstrated~\cite{aghaeerad2025, jia2025, wang2025a, jia2026}.\\    
    \begin{figure*}[t]
        \includegraphics[width=\textwidth]{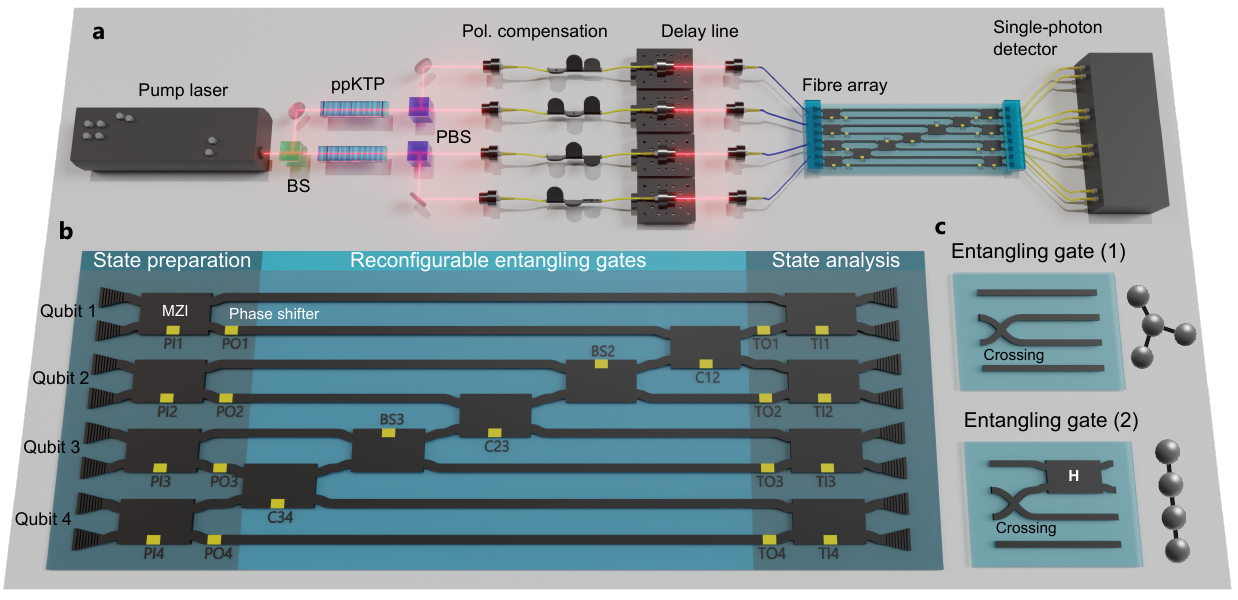}
        \caption{Experimental setup. 
        \textbf{a} Two type-II \gls{spdc} sources generate single photons at \qty{1550}{\nano\metre} wavelength. 
        After temporal synchronisation using delay lines, the photons are coupled into the \acrlong{pic} via a polarisation-maintaining fibre array and on-chip grating couplers, 
        a second fibre array routes the photons to \acrfullpl{snspd}. 
        The circuit is an 8-mode reconfigurable \acrlong{soi} chip comprising \acrfullpl{mmi} as balanced beam splitters and thermo-optic phase shifters.  
        \textbf{b} 
        Zoom-in of the chip, where yellow rectangles indicate phase shifters whilst black boxes denote \acrfullpl{mzi}. These are based on two \glspl{mmi} and a thermo-optic phase shifter.
        The chip consists of three sections: state preparation, reconfigurable entangling gates, and state analysis. 
        \textbf{c} 
        The concatenation of the entangling gates $(n-1)$ times creates an $n$-qubit star graph state in configuration (1), and an $n$-qubit linear graph state in configuration (2). 
        In both configurations, the \glspl{mzi} between different qubits act as a waveguide crossing, 
        whilst the \gls{mzi} acting on a qubit implements the identity in configuration (1) and a Hadamard gate ($H$) in configuration (2). 
         \label{fig:setup} } 
    \end{figure*} 
    Here, we demonstrate \gls{mbqc} on a \acrlong{pic} by generating and manipulating four-qubit entangled resource states, including linear cluster states and star graph states. 
    These resource states enable the implementation of a range of \gls{mbqc} protocols, including a non-Clifford single-qubit gate, two-qubit gates, and demonstrations of Grover's search algorithm~\cite{grover1996, grover1997} and the Deutsch–Jozsa algorithm~\cite{deutsch1992}. 
    Our results show that integrated photonics is capable of supporting reconfigurable four-photon \gls{mbqc}, providing a foundation for future larger-scale implementations.    
   
\section{Theoretical background}
    \Gls{mbqc} implements quantum gates through adaptive single-qubit measurements on graph states.
    These states are represented by vertices, i.e. qubits initialised in the state $\state{\p}$, where 
    $\state{+}=\frac{1}{\sqrt{2}}(\state{0} + \state{1})$, and edges that indicate entanglement generated via controlled-$Z$ (C$Z$) gates.
    Large-scale \gls{mbqc} typically uses cluster states, i.e. graph states arranged in a rectangular lattice~\cite{raussendorf2001, briegel2009}. 
    In a two-dimensional cluster state, logical information is processed through successive single-qubit measurements. The rows encode the logical qubits, while the columns correspond to the computational depth.\\
    We illustrate the working principle of \gls{mbqc} by considering the example of a linear cluster state consisting of four qubits.  
    Measuring the first three qubits in the basis $B(\alpha_1)$, $B(\alpha_2)$ and $B(\alpha_3)$, where 
    $B(\alpha) = \{\state{+_\alpha}, \state{-_\alpha} \}$ with 
    $ \state{\pm_\alpha} = (\state{0} \pm  \exp(i\alpha)\state{1})/\sqrt{2}$, projects the last qubit into the state 
    \begin{equation}
        \begin{split}
        \state{\psi_4} =& X^{m_3}Z^{m_2}X^{m_1}H \label{eq:mbqc_1} \\
                            &R_z((-1)^{m_2}\alpha_3)R_x((-1)^{m_1}\alpha_2)
                            R_z(\alpha_1)\state{+},
        \end{split}
    \end{equation}    
    where $m_i$ are the measurement outcomes. 
    \begin{figure*}[t]
    \includegraphics[width=\textwidth]{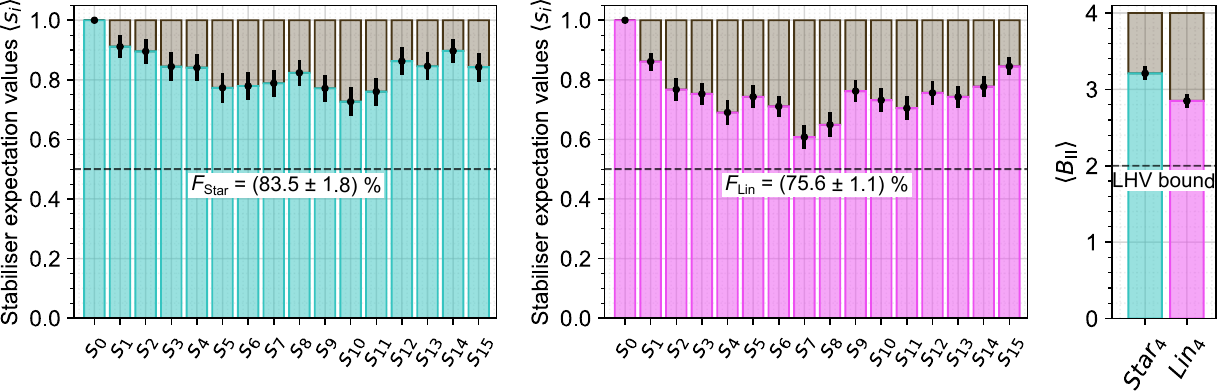}
    \caption{Graph state characterisation via stabiliser expectation values. 
        \textbf{a, b} Stabiliser expectation values for the generated 
        $\state{\text{Star}_4}$ and $\state{\text{Lin}_4}$ state. 
        The dashed line marks the \qty{50}{\percent} threshold for genuine multi-partite entanglement.
        \textbf{c} Verification of non-locality by violation of a two-setting 
        Mermin-type Bell inequality, with $\left|\exval{B_{\text{II}}}_{\text{Star}}\right|=\qty{3.21 \pm 0.09 }{}$ and $\left|\exval{B_{\text{II}}}_{\text{Lin}}\right|=\qty{2.85 \pm 0.09 }{}$.
        \label{fig:stabilisers}}
    \end{figure*}
    This measurement sequence implements an arbitrary single-qubit gate.         
    Two-qubit gates are realised by leveraging the cluster states entanglement structure~\cite{raussendorf2001, briegel2009}. 
    Universal \gls{mbqc} needs feed-forward operations to adapt the measurement settings based on the previous measurement outcomes $m_i$ (see Eq. \eqref{eq:mbqc_1}). 
    Here, we demonstrate \gls{mbqc} based on four-qubit graph states (see Fig.~\ref{fig:stabilisers} and Appendix):
    \begin{align}
        \state{\text{Lin}_4} &=CZ_{12}CZ_{23}CZ_{34}\state{\p\p\p\p}, \\
        \state{\text{Star}_4} &=CZ_{12}CZ_{13}CZ_{14}\state{\p\p\p\p}. 
    \end{align}

\section{Experiment}
    We implement \gls{mbqc} on a \gls{soi} photonic chip that encodes four photonic qubits in dual-rail encoding across eight modes.
    Tuneable beamsplitters and phase shifters are utilised to implement postselected entangling
    gates and set the measurement bases (see Fig.~\ref{fig:setup} b). 
    We generate single photons at \qty{1550}{\nano\metre} from two type-II \gls{spdc} sources, 
    each based on a \gls{ppktp} crystal pumped by a picosecond \qty{775}{\nano\metre} Ti:Sa laser at a repetition rate of $\qty{76}{\mega\hertz}$.
    We couple the photons into the chip through polarisation-maintaining fibres aligned to on-chip grating couplers. After the on-chip operations, the photons are coupled out via a second set of grating couplers and routed through single-mode fibres with \qty{1550}{\nano\metre} bandpass filters to eight \glspl{snspd}.
    We correct for detection-efficiency differences by normalizing the data with independently measured detector efficiencies.

\section{Results}
    \subsection{Graph state generation} 
        We create four-qubit states locally equivalent to $\state{\text{Star}_4}$ and linear graph states $\state{\text{Lin}_4}$ from single photon inputs with a success probability of $1/2^{3}$~\cite{dhand2018, meyer-scott2022}.
        As graph states are stabiliser states, their fidelity can be estimated from stabiliser measurements.
        A direct measurement of the fidelity $F = \text{mean}\left( \exval{s_i} \right)$ can be obtained  from at most $2^n$ measurement settings~\cite{guhne2009}, where $s_i$ are the stabiliser and $\exval{...}$ denotes an expectation value (see Fig.~\ref{fig:stabilisers}).
        We obtain a direct fidelity estimation of $F_{\text{Star}_4}=\qty{83.5\pm 1.8}{\percent}$      
        and $F_{\text{Lin}_4}=\qty{75.6 \pm 1.1}{\percent}$ at rates of $\qty{12.1}{\milli\hertz}$ and $\qty{13.3}{\milli\hertz}$, respectively. 
        The measurement settings for each stabiliser can be found in the Appendix.\\  
        Both generated graph states surpass the $\qty{50}{\percent}$ fidelity threshold and are therefore genuinely multipartite entangled~\cite{toth2005}. 
        Additionally, we verify the nonlocality of the states by the violation of a Mermin-type Bell inequality.
        \begin{figure*}[t]
            \includegraphics[width=\textwidth]{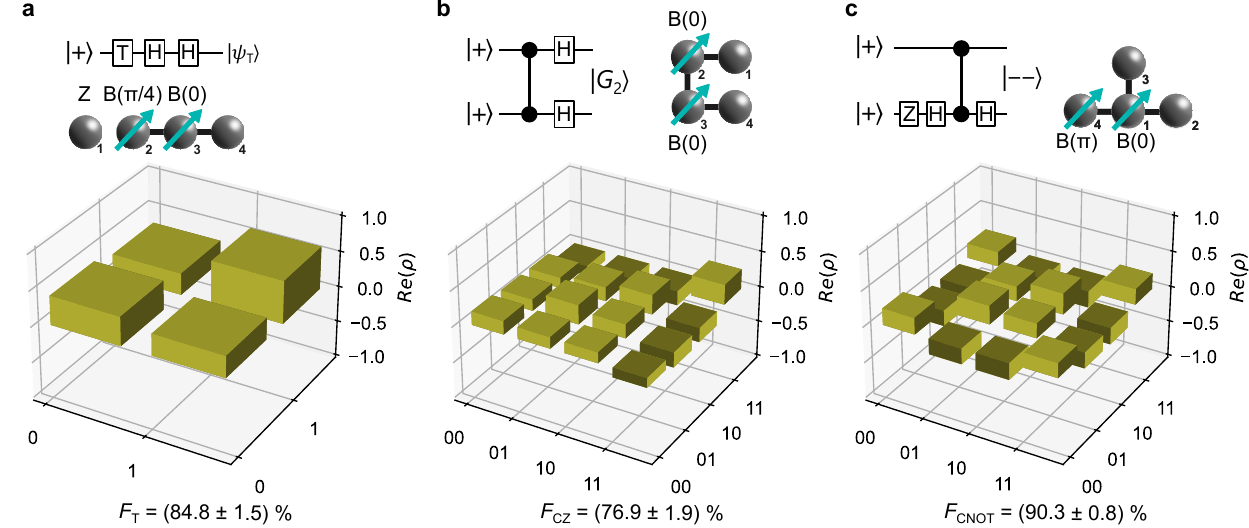}
            \caption{A set of implemented quantum gates, their implementation via MBQC and the resulting output states. The measurement bases are denoted by  $B(\alpha) = \{\state{+_\alpha}, \state{-_\alpha} \}$ and $Z=\{\state{0}, \state{1} \}$.
            \textbf{a}, \textbf{b} 
            T-gate operation $T\state{\p}=\frac{1}{\sqrt{2}} (\state{0}+\exp(\frac{i\pi}{4})\state{1})$
            and controlled-Z operation $CZ \state{\p\p}$ on a four-qubit linear graph state.
            \textbf{c} 
            CNOT operation $\text{CNOT} \state{\p\m} = \state{\m\m}$ implemented on star graph state.
            \label{fig:mbqc}
            } 
        \end{figure*}
        The bounds are given by the maximum expectation value that can be obtained from a local hidden-variable (LHV) description.
        We test a two-setting Bell inequality $|\exval{\inequal_{\text{II}}}|\leq 2$ constructed from a subset of the stabilisers to optimise the violation threshold~\cite{cabello2008}.
        The results violate the inequality, confirming the nonlocality of both generated states (see Fig.~\ref{fig:stabilisers}~c).

    \subsection{Measurement-based quantum computing}  
        A universal set of gate operations is a key requirement for quantum computation. We implement both single- and two-qubit \gls{mbqc} gates, thereby demonstrating the fundamental gate primitives required for universal quantum computation (see Fig.~\ref{fig:mbqc}).
        As the required non-Clifford operation for universal quantum computation, we implement the $T$ gate with an output-state fidelity of $F_T = \qty{84.8 \pm 1.5}{\percent}$, obtained from a full state tomography. 
        The density matrix was reconstructed from a \gls{mle} on the measured data (see Fig.~\ref{fig:mbqc}~a).\\
        Additionally, we realise two-qubit operations by implementing  a controlled-Z gate (CZ) and a controlled-NOT (CNOT) gate, yielding output state fidelities of $F_{\text{C}Z} =\qty{76.9 \pm 1.9}{\percent}$ and $F_{\text{CNOT}} = \qty{90.3 \pm 0.8}{\percent}$, respectively (see Fig.~\ref{fig:mbqc}~\textbf{b}~and~\textbf{c}).          
        
        \subsubsection{Quantum algorithms}\label{sec:quantum_algos}
\begin{figure*}[t]
            \includegraphics[width=\textwidth]{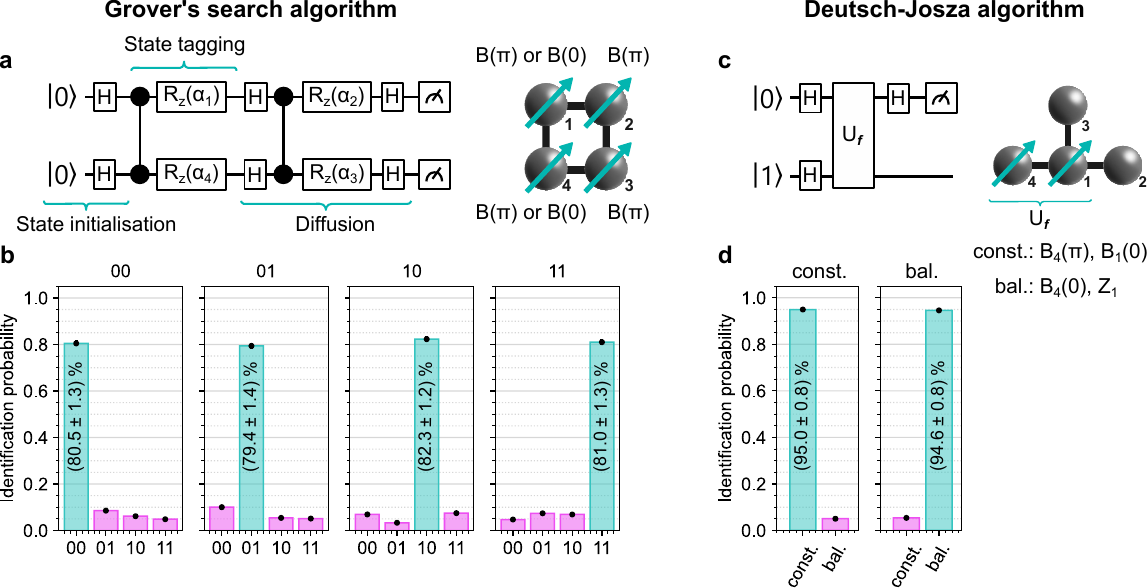}
            \caption{\gls{mbqc} implementation of Grover's search and Deutsch-Josza algorithm. 
                \textbf{a} Circuit representation of the two-qubit Grover's search algorithm and 
                    the four-qubit box graph state used for the \gls{mbqc} implementation.
                \textbf{b} Measured identification probabilities for Grover's algorithm (expected results in cyan).
                \textbf{c} 
                    Circuit and \gls{mbqc} implementation for a one-bit Deutsch-Josza algorithm 
                    using a four-qubit star graph state.
                \textbf{d} Identification probabilities for the constant function and the balanced function.
                    Expected outcomes in cyan. 
                \label{fig:algos} } 
        \end{figure*}

            We demonstrate Grover's search algorithm on a four-element search space. 
            The algorithm consists of a tagging operation, followed by an inversion about the mean and a final measurement to reveal the tagged element. 
            Its \gls{mbqc} implementation is based on measurements on a four-qubit box state \cite{walther2005}.
            Measuring qubits 1 and 4 implements the tagging and inversion, whilst measurements on qubits 2 and 3 retrieve the outcome of the algorithm (see Fig.~\ref{fig:algos}~a). 
            We perform the algorithm for all four elements obtaining an average identification probability of 
            $\qty{80.8 \pm 0.7}{\percent}$ (see Fig.~\ref{fig:algos}~\textbf{b}).\\       
            We further demonstrate an on-chip implementation of the Deutsch-Jozsa algorithm, 
            which determines whether a function $f:{0,1}\rightarrow{0,1}$ is constant or balanced, returning $0$ for half the inputs and $1$ for the other half.
            The \gls{mbqc}-based implementation of the algorithm is shown in Fig.~\ref{fig:algos}~\textbf{c}. We demonstrate the algorithm using the constant function $f(0)=f(1)=0$ and the balanced function $f(0)=0, f(1)=1$, realised by the oracles $U_f=\mathds{1}\otimes\mathds{1}$ and $U_f=\mathrm{CNOT}$, respectively. Correct classification is achieved with probabilities of $\qty{95.0+-0.8}{\percent}$ for the constant function and $\qty{94.6 +- 0.8}{\percent}$ for the balanced function (see Fig.~\ref{fig:algos}~\textbf{d}).

\section{Summary and discussion}
    In this work we demonstrated \acrfull{mbqc} on a photonic integrated circuit.
    We implemented the Grover's search algorithm and the Deutsch-Josza algorithm, with average identification probabilities above $\qty{80}{\percent}$ and $\qty{94}{\percent}$, respectively. 
    We achieved an on-chip four-photon linear graph state fidelity of $F_\text{Lin}=\qty{75.6 \pm 1.1}{\percent}$, 
    surpassing, to our knowledge, the highest reported on-chip fidelity.
    Additionally, we obtained a four-photon star graph state with $F_\text{Star}=\qty{83.5\pm 1.8}{\percent}$, 
    comparable to the current on-chip state-of-the-art.
    The high fidelities were enabled by high fibre-to-chip coupling efficiencies and deep-trench thermo-optic 
    phase shifters, which effectively reduce thermal crosstalk.
    The main source of error is attributed to higher-order noise in combination with losses and subsequently 
    non-perfect photon indistinguishability, path-dependent losses and phase shifter drifts.\\    
    To overcome these limitations in the future, the chip design is compatible with deterministic single-photon sources at $\qty{1550}{\nano\metre}$ wavelength, 
    which are currently seeing rapid development~\cite{Kim2025, Pfister2025, Hauser2026}.
    These could enable the generation of larger graph states, as well as heralded generation 
    schemes~\cite{PsiQSwitch2021, chen2024, Cao2024, maring2024} and heralded fusion operations~\cite{Browne2005}, 
    as building blocks for fusion-based quantum computing~\cite{PsiQFusion2023}.
    The SOI platform offers high integration density and low-loss waveguides, making it well suited for large-scale photonic circuits. Active feed-forward, however, benefits from complementary technologies that provide high-speed, low-loss switching and ultra-low-loss optical delays.
    A promising route forward is the interfacing of multiple materials to gain access to ultra-low loss propagation 
    in combination with fast-switching capabilities~\cite{PsiQComponents2025}, providing a path toward large-scale photonic \gls{mbqc}.
\section{Acknowledgements}
    We thank Markus Greul (IMS CHIPS) for overseeing the chip fabrication and Stephan Schmiel (IMS CHIPS) for wire bonding the PIC. 
    We also thank Shiang-Yu Huang for designing the grating couplers used in this work. 
    Furthermore, we thank Matthias J. Bayerbach and Nico Hauser for technical support with the \glspl{snspd} and the pump laser, respectively, as well as Simone E. D'Aurelio for helpful discussions regarding data acquisition. Finally, we thank Daniel Bhatti for valuable insights into graph state generation.\\
    We acknowledge the support from 
    the Federal Ministry of Research, Technology and Space (BMFTR, projects
    SiSiQ: FKZ 13N14920, 
    PhotonQ: FKZ 13N15758, 
    QR.N: FKZ 16KIS2207, 
    TD.QR: FKZ 16KISS026, 
    QCyber: 16KIS2590K), 
    and the Deutsche Forschungsgemeinschaft (DFG, German Research Foundation, 431314977/GRK2642, 516238647/SFB 1667, 563437379/SPP2514).
    the Carl Zeiss Foundation, and 
    the Centre for Integrated Quantum Science and Technology (IQST).

\bibliographystyle{apsrev4-2-truncated}
\bibliography{short2} 

\begin{thebibliography}{51}%
\makeatletter
\providecommand \@ifxundefined [1]{%
 \@ifx{#1\undefined}
}%
\providecommand \@ifnum [1]{%
 \ifnum #1\expandafter \@firstoftwo
 \else \expandafter \@secondoftwo
 \fi
}%
\providecommand \@ifx [1]{%
 \ifx #1\expandafter \@firstoftwo
 \else \expandafter \@secondoftwo
 \fi
}%
\providecommand \natexlab [1]{#1}%
\providecommand \enquote  [1]{``#1''}%
\providecommand \bibnamefont  [1]{#1}%
\providecommand \bibfnamefont [1]{#1}%
\providecommand \citenamefont [1]{#1}%
\providecommand \href@noop [0]{\@secondoftwo}%
\providecommand \href [0]{\begingroup \@sanitize@url \@href}%
\providecommand \@href[1]{\@@startlink{#1}\@@href}%
\providecommand \@@href[1]{\endgroup#1\@@endlink}%
\providecommand \@sanitize@url [0]{\catcode `\\12\catcode `\$12\catcode `\&12\catcode `\#12\catcode `\^12\catcode `\_12\catcode `\%12\relax}%
\providecommand \@@startlink[1]{}%
\providecommand \@@endlink[0]{}%
\providecommand \url  [0]{\begingroup\@sanitize@url \@url }%
\providecommand \@url [1]{\endgroup\@href {#1}{\urlprefix }}%
\providecommand \urlprefix  [0]{URL }%
\providecommand \Eprint [0]{\href }%
\providecommand \doibase [0]{https://doi.org/}%
\providecommand \selectlanguage [0]{\@gobble}%
\providecommand \bibinfo  [0]{\@secondoftwo}%
\providecommand \bibfield  [0]{\@secondoftwo}%
\providecommand \translation [1]{[#1]}%
\providecommand \BibitemOpen [0]{}%
\providecommand \bibitemStop [0]{}%
\providecommand \bibitemNoStop [0]{.\EOS\space}%
\providecommand \EOS [0]{\spacefactor3000\relax}%
\providecommand \BibitemShut  [1]{\csname bibitem#1\endcsname}%
\let\auto@bib@innerbib\@empty
\bibitem [{\citenamefont {Couteau}\ \emph {et~al.}(2023)\citenamefont {Couteau}, \citenamefont {Barz}, \citenamefont {Durt}, \citenamefont {Gerrits}, \citenamefont {Huwer}, \citenamefont {Prevedel}, \citenamefont {Rarity}, \citenamefont {Shields},\ and\ \citenamefont {Weihs}}]{couteau2023}%
  \BibitemOpen
  \bibfield  {author} {\bibinfo {author} {\bibfnamefont {C.}~\bibnamefont {Couteau}}, \bibinfo {author} {\bibfnamefont {S.}~\bibnamefont {Barz}}, \bibinfo {author} {\bibfnamefont {T.}~\bibnamefont {Durt}}, \bibinfo {author} {\bibfnamefont {T.}~\bibnamefont {Gerrits}}, \bibinfo {author} {\bibfnamefont {J.}~\bibnamefont {Huwer}}, \emph {et~al.},\ }\href {https://doi.org/10.1038/s42254-023-00583-2} {\bibfield  {journal} {\bibinfo  {journal} {Nature Reviews Physics}\ }\textbf {\bibinfo {volume} {5}},\ \bibinfo {pages} {326} (\bibinfo {year} {2023})}\BibitemShut {NoStop}%
\bibitem [{\citenamefont {Raussendorf}\ and\ \citenamefont {Briegel}(2001)}]{raussendorf2001}%
  \BibitemOpen
  \bibfield  {author} {\bibinfo {author} {\bibfnamefont {R.}~\bibnamefont {Raussendorf}}\ and\ \bibinfo {author} {\bibfnamefont {H.~J.}\ \bibnamefont {Briegel}},\ }\href {https://doi.org/10.1103/PhysRevLett.86.5188} {\bibfield  {journal} {\bibinfo  {journal} {Physical Review Letters}\ }\textbf {\bibinfo {volume} {86}},\ \bibinfo {pages} {5188} (\bibinfo {year} {2001})}\BibitemShut {NoStop}%
\bibitem [{\citenamefont {Briegel}\ \emph {et~al.}(2009)\citenamefont {Briegel}, \citenamefont {Browne}, \citenamefont {D{\"u}r}, \citenamefont {Raussendorf},\ and\ \citenamefont {{Van den Nest}}}]{briegel2009}%
  \BibitemOpen
  \bibfield  {author} {\bibinfo {author} {\bibfnamefont {H.~J.}\ \bibnamefont {Briegel}}, \bibinfo {author} {\bibfnamefont {D.~E.}\ \bibnamefont {Browne}}, \bibinfo {author} {\bibfnamefont {W.}~\bibnamefont {D{\"u}r}}, \bibinfo {author} {\bibfnamefont {R.}~\bibnamefont {Raussendorf}},\ and\ \bibinfo {author} {\bibfnamefont {M.}~\bibnamefont {{Van den Nest}}},\ }\href {https://doi.org/10.1038/nphys1157} {\bibfield  {journal} {\bibinfo  {journal} {Nature Physics}\ }\textbf {\bibinfo {volume} {5}},\ \bibinfo {pages} {19} (\bibinfo {year} {2009})}\BibitemShut {NoStop}%
\bibitem [{\citenamefont {Browne}\ and\ \citenamefont {Rudolph}(2005)}]{Browne2005}%
  \BibitemOpen
  \bibfield  {author} {\bibinfo {author} {\bibfnamefont {D.~E.}\ \bibnamefont {Browne}}\ and\ \bibinfo {author} {\bibfnamefont {T.}~\bibnamefont {Rudolph}},\ }\href {https://doi.org/10.1103/PhysRevLett.95.010501} {\bibfield  {journal} {\bibinfo  {journal} {Physical Review Letters}\ }\textbf {\bibinfo {volume} {95}},\ \bibinfo {pages} {010501} (\bibinfo {year} {2005})}\BibitemShut {NoStop}%
\bibitem [{\citenamefont {Bartolucci}\ \emph {et~al.}(2023)\citenamefont {Bartolucci}, \citenamefont {Birchall}, \citenamefont {Bomb{\'i}n}, \citenamefont {Cable}, \citenamefont {Dawson}, \citenamefont {{Gimeno-Segovia}}, \citenamefont {Johnston}, \citenamefont {Kieling}, \citenamefont {Nickerson}, \citenamefont {Pant}, \citenamefont {Pastawski}, \citenamefont {Rudolph},\ and\ \citenamefont {Sparrow}}]{PsiQFusion2023}%
  \BibitemOpen
  \bibfield  {author} {\bibinfo {author} {\bibfnamefont {S.}~\bibnamefont {Bartolucci}}, \bibinfo {author} {\bibfnamefont {P.}~\bibnamefont {Birchall}}, \bibinfo {author} {\bibfnamefont {H.}~\bibnamefont {Bomb{\'i}n}}, \bibinfo {author} {\bibfnamefont {H.}~\bibnamefont {Cable}}, \bibinfo {author} {\bibfnamefont {C.}~\bibnamefont {Dawson}}, \emph {et~al.},\ }\href {https://doi.org/10.1038/s41467-023-36493-1} {\bibfield  {journal} {\bibinfo  {journal} {Nature Communications}\ }\textbf {\bibinfo {volume} {14}},\ \bibinfo {pages} {1} (\bibinfo {year} {2023})},\ \Eprint {https://arxiv.org/abs/2101.09310} {arXiv:2101.09310} \BibitemShut {NoStop}%
\bibitem [{\citenamefont {Walther}\ \emph {et~al.}(2005)\citenamefont {Walther}, \citenamefont {Resch}, \citenamefont {Rudolph}, \citenamefont {Schenck}, \citenamefont {Weinfurter}, \citenamefont {Vedral}, \citenamefont {Aspelmeyer},\ and\ \citenamefont {Zeilinger}}]{walther2005}%
  \BibitemOpen
  \bibfield  {author} {\bibinfo {author} {\bibfnamefont {P.}~\bibnamefont {Walther}}, \bibinfo {author} {\bibfnamefont {K.~J.}\ \bibnamefont {Resch}}, \bibinfo {author} {\bibfnamefont {T.}~\bibnamefont {Rudolph}}, \bibinfo {author} {\bibfnamefont {E.}~\bibnamefont {Schenck}}, \bibinfo {author} {\bibfnamefont {H.}~\bibnamefont {Weinfurter}}, \emph {et~al.},\ }\href {https://doi.org/10.1038/nature03347} {\bibfield  {journal} {\bibinfo  {journal} {Nature}\ }\textbf {\bibinfo {volume} {434}},\ \bibinfo {pages} {169} (\bibinfo {year} {2005})}\BibitemShut {NoStop}%
\bibitem [{\citenamefont {Prevedel}\ \emph {et~al.}(2007)\citenamefont {Prevedel}, \citenamefont {Walther}, \citenamefont {Tiefenbacher}, \citenamefont {B{\"o}hi}, \citenamefont {Kaltenbaek}, \citenamefont {Jennewein},\ and\ \citenamefont {Zeilinger}}]{prevedel2007}%
  \BibitemOpen
  \bibfield  {author} {\bibinfo {author} {\bibfnamefont {R.}~\bibnamefont {Prevedel}}, \bibinfo {author} {\bibfnamefont {P.}~\bibnamefont {Walther}}, \bibinfo {author} {\bibfnamefont {F.}~\bibnamefont {Tiefenbacher}}, \bibinfo {author} {\bibfnamefont {P.}~\bibnamefont {B{\"o}hi}}, \bibinfo {author} {\bibfnamefont {R.}~\bibnamefont {Kaltenbaek}}, \emph {et~al.},\ }\href {https://doi.org/10.1038/nature05346} {\bibfield  {journal} {\bibinfo  {journal} {Nature}\ }\textbf {\bibinfo {volume} {445}},\ \bibinfo {pages} {65} (\bibinfo {year} {2007})}\BibitemShut {NoStop}%
\bibitem [{\citenamefont {Tame}\ \emph {et~al.}(2007)\citenamefont {Tame}, \citenamefont {Prevedel}, \citenamefont {Paternostro}, \citenamefont {B{\"o}hi}, \citenamefont {Kim},\ and\ \citenamefont {Zeilinger}}]{tame2007}%
  \BibitemOpen
  \bibfield  {author} {\bibinfo {author} {\bibfnamefont {M.~S.}\ \bibnamefont {Tame}}, \bibinfo {author} {\bibfnamefont {R.}~\bibnamefont {Prevedel}}, \bibinfo {author} {\bibfnamefont {M.}~\bibnamefont {Paternostro}}, \bibinfo {author} {\bibfnamefont {P.}~\bibnamefont {B{\"o}hi}}, \bibinfo {author} {\bibfnamefont {M.~S.}\ \bibnamefont {Kim}}, \emph {et~al.},\ }\href {https://doi.org/10.1103/PhysRevLett.98.140501} {\bibfield  {journal} {\bibinfo  {journal} {Physical Review Letters}\ }\textbf {\bibinfo {volume} {98}},\ \bibinfo {pages} {140501} (\bibinfo {year} {2007})}\BibitemShut {NoStop}%
\bibitem [{\citenamefont {Lee}\ \emph {et~al.}(2012)\citenamefont {Lee}, \citenamefont {Park}, \citenamefont {Cho}, \citenamefont {Kang}, \citenamefont {Lee}, \citenamefont {Kim}, \citenamefont {Lee},\ and\ \citenamefont {Choi}}]{lee2012}%
  \BibitemOpen
  \bibfield  {author} {\bibinfo {author} {\bibfnamefont {S.~M.}\ \bibnamefont {Lee}}, \bibinfo {author} {\bibfnamefont {H.~S.}\ \bibnamefont {Park}}, \bibinfo {author} {\bibfnamefont {J.}~\bibnamefont {Cho}}, \bibinfo {author} {\bibfnamefont {Y.}~\bibnamefont {Kang}}, \bibinfo {author} {\bibfnamefont {J.~Y.}\ \bibnamefont {Lee}}, \emph {et~al.},\ }\href {https://doi.org/10.1364/OE.20.006915} {\bibfield  {journal} {\bibinfo  {journal} {Optics Express}\ }\textbf {\bibinfo {volume} {20}},\ \bibinfo {pages} {6915} (\bibinfo {year} {2012})}\BibitemShut {NoStop}%
\bibitem [{\citenamefont {Barz}\ \emph {et~al.}(2012)\citenamefont {Barz}, \citenamefont {Kashefi}, \citenamefont {Broadbent}, \citenamefont {Fitzsimons}, \citenamefont {Zeilinger},\ and\ \citenamefont {Walther}}]{Barz2012}%
  \BibitemOpen
  \bibfield  {author} {\bibinfo {author} {\bibfnamefont {S.}~\bibnamefont {Barz}}, \bibinfo {author} {\bibfnamefont {E.}~\bibnamefont {Kashefi}}, \bibinfo {author} {\bibfnamefont {A.}~\bibnamefont {Broadbent}}, \bibinfo {author} {\bibfnamefont {J.~F.}\ \bibnamefont {Fitzsimons}}, \bibinfo {author} {\bibfnamefont {A.}~\bibnamefont {Zeilinger}}, \emph {et~al.},\ }\href {https://doi.org/10.1126/science.1214707} {\bibfield  {journal} {\bibinfo  {journal} {Science}\ }\textbf {\bibinfo {volume} {335}},\ \bibinfo {pages} {303} (\bibinfo {year} {2012})}\BibitemShut {NoStop}%
\bibitem [{\citenamefont {Vallone}\ \emph {et~al.}(2008)\citenamefont {Vallone}, \citenamefont {Pomarico}, \citenamefont {De~Martini},\ and\ \citenamefont {Mataloni}}]{vallone2008}%
  \BibitemOpen
  \bibfield  {author} {\bibinfo {author} {\bibfnamefont {G.}~\bibnamefont {Vallone}}, \bibinfo {author} {\bibfnamefont {E.}~\bibnamefont {Pomarico}}, \bibinfo {author} {\bibfnamefont {F.}~\bibnamefont {De~Martini}},\ and\ \bibinfo {author} {\bibfnamefont {P.}~\bibnamefont {Mataloni}},\ }\href {https://doi.org/10.1103/PhysRevLett.100.160502} {\bibfield  {journal} {\bibinfo  {journal} {Physical Review Letters}\ }\textbf {\bibinfo {volume} {100}},\ \bibinfo {pages} {160502} (\bibinfo {year} {2008})}\BibitemShut {NoStop}%
\bibitem [{\citenamefont {Zhu}\ \emph {et~al.}(2026)\citenamefont {Zhu}, \citenamefont {Chen}, \citenamefont {Ma}, \citenamefont {Zhao}, \citenamefont {Guo}, \citenamefont {Liang}, \citenamefont {Tu}, \citenamefont {He}, \citenamefont {Luo}, \citenamefont {Wang}, \citenamefont {Song}, \citenamefont {Jin}, \citenamefont {Liu}, \citenamefont {Yang}, \citenamefont {Wang}, \citenamefont {Leong}, \citenamefont {Wang},\ and\ \citenamefont {Yang}}]{zhu2026}%
  \BibitemOpen
  \bibfield  {author} {\bibinfo {author} {\bibfnamefont {H.}~\bibnamefont {Zhu}}, \bibinfo {author} {\bibfnamefont {T.}~\bibnamefont {Chen}}, \bibinfo {author} {\bibfnamefont {H.}~\bibnamefont {Ma}}, \bibinfo {author} {\bibfnamefont {Z.}~\bibnamefont {Zhao}}, \bibinfo {author} {\bibfnamefont {J.}~\bibnamefont {Guo}}, \emph {et~al.},\ }\href {https://doi.org/10.1038/s44310-026-00114-8} {\bibfield  {journal} {\bibinfo  {journal} {npj Nanophotonics}\ }\textbf {\bibinfo {volume} {3}},\ \bibinfo {pages} {20} (\bibinfo {year} {2026})}\BibitemShut {NoStop}%
\bibitem [{\citenamefont {Alexander}\ \emph {et~al.}(2025)\citenamefont {Alexander}, \citenamefont {Benyamini}, \citenamefont {Black}, \citenamefont {Bonneau}, \citenamefont {Burgos}, \citenamefont {Burridge}, \citenamefont {Cable}, \citenamefont {Campbell}, \citenamefont {Catalano}, \citenamefont {Ceballos}, \citenamefont {Chang}, \citenamefont {Choudhury}, \citenamefont {Chung}, \citenamefont {Danesh}, \citenamefont {Dauer}, \citenamefont {Davis}, \citenamefont {Dudley}, \citenamefont {{Er-Xuan}}, \citenamefont {Fargas}, \citenamefont {Farsi}, \citenamefont {Fenrich}, \citenamefont {Frazer}, \citenamefont {Fukami}, \citenamefont {Ganesan}, \citenamefont {Gibson}, \citenamefont {{Gimeno-Segovia}}, \citenamefont {Goeldi}, \citenamefont {Goley}, \citenamefont {Haislmaier}, \citenamefont {Halimi}, \citenamefont {Hansen}, \citenamefont {Hardy}, \citenamefont {Horng}, \citenamefont {House}, \citenamefont {Hu}, \citenamefont {Jadidi}, \citenamefont {Jain}, \citenamefont {Johansson}, \citenamefont {Jones}, \citenamefont {Kamineni}, \citenamefont {Kelez}, \citenamefont {Koustuban}, \citenamefont {Kovall}, \citenamefont {Krogen}, \citenamefont {Kumar}, \citenamefont {Liang}, \citenamefont {LiCausi}, \citenamefont {Llewellyn}, \citenamefont {Lokovic}, \citenamefont {Lovelady}, \citenamefont {Manfrinato}, \citenamefont {Melnichuk}, \citenamefont {Mendoza}, \citenamefont {Moores}, \citenamefont {Mukherjee}, \citenamefont {Munns}, \citenamefont {Musalem}, \citenamefont {Najafi}, \citenamefont {O'Brien}, \citenamefont {Ortmann}, \citenamefont {Pai}, \citenamefont {Park}, \citenamefont {Peng}, \citenamefont {Penthorn}, \citenamefont {Peterson}, \citenamefont {Peterson}, \citenamefont {Poush}, \citenamefont {Pryde}, \citenamefont {Ramprasad}, \citenamefont {Ray}, \citenamefont {Rodriguez}, \citenamefont {Roxworthy}, \citenamefont {Rudolph}, \citenamefont {Saunders}, \citenamefont {Shadbolt}, \citenamefont {Shah}, \citenamefont {Bahgat~Shehata}, \citenamefont {Shin}, \citenamefont {Sinsky}, \citenamefont {Smith}, \citenamefont {Sohn}, \citenamefont {Sohn}, \citenamefont {Son}, \citenamefont {Souza}, \citenamefont {Sparrow}, \citenamefont {Staffaroni}, \citenamefont {Stavrakas}, \citenamefont {Sukumaran}, \citenamefont {Tamborini}, \citenamefont {Thompson}, \citenamefont {Tran}, \citenamefont {Triplett}, \citenamefont {Tung}, \citenamefont {Veitia}, \citenamefont {Vert}, \citenamefont {Vidrighin}, \citenamefont {Vorobeichik}, \citenamefont {Weigel}, \citenamefont {Wingert}, \citenamefont {Wooding},\ and\ \citenamefont {Zhou}}]{PsiQComponents2025}%
  \BibitemOpen
  \bibfield  {author} {\bibinfo {author} {\bibfnamefont {K.}~\bibnamefont {Alexander}}, \bibinfo {author} {\bibfnamefont {A.}~\bibnamefont {Benyamini}}, \bibinfo {author} {\bibfnamefont {D.}~\bibnamefont {Black}}, \bibinfo {author} {\bibfnamefont {D.}~\bibnamefont {Bonneau}}, \bibinfo {author} {\bibfnamefont {S.}~\bibnamefont {Burgos}}, \emph {et~al.},\ }\href {https://doi.org/10.1038/s41586-025-08820-7} {\bibfield  {journal} {\bibinfo  {journal} {Nature}\ }\textbf {\bibinfo {volume} {641}},\ \bibinfo {pages} {876} (\bibinfo {year} {2025})}\BibitemShut {NoStop}%
\bibitem [{\citenamefont {Sund}\ \emph {et~al.}(2023)\citenamefont {Sund}, \citenamefont {Lomonte}, \citenamefont {Paesani}, \citenamefont {Wang}, \citenamefont {Carolan}, \citenamefont {Bart}, \citenamefont {Wieck}, \citenamefont {Ludwig}, \citenamefont {Midolo}, \citenamefont {Pernice}, \citenamefont {Lodahl},\ and\ \citenamefont {Lenzini}}]{sund2023}%
  \BibitemOpen
  \bibfield  {author} {\bibinfo {author} {\bibfnamefont {P.~I.}\ \bibnamefont {Sund}}, \bibinfo {author} {\bibfnamefont {E.}~\bibnamefont {Lomonte}}, \bibinfo {author} {\bibfnamefont {S.}~\bibnamefont {Paesani}}, \bibinfo {author} {\bibfnamefont {Y.}~\bibnamefont {Wang}}, \bibinfo {author} {\bibfnamefont {J.}~\bibnamefont {Carolan}}, \emph {et~al.},\ }\bibfield  {journal} {\bibinfo  {journal} {Science Advances}\ }\textbf {\bibinfo {volume} {9}},\ \href {https://doi.org/10.1126/sciadv.adg7268} {10.1126/sciadv.adg7268} (\bibinfo {year} {2023})\BibitemShut {NoStop}%
\bibitem [{\citenamefont {Zhang}\ \emph {et~al.}(2026)\citenamefont {Zhang}, \citenamefont {Lau}, \citenamefont {Li}, \citenamefont {Foo}, \citenamefont {Yoo}, \citenamefont {Pan}, \citenamefont {Tan}, \citenamefont {Singh}, \citenamefont {Lim}, \citenamefont {Tobing}, \citenamefont {Luo},\ and\ \citenamefont {Yeo}}]{zhang2026}%
  \BibitemOpen
  \bibfield  {author} {\bibinfo {author} {\bibfnamefont {Y.}~\bibnamefont {Zhang}}, \bibinfo {author} {\bibfnamefont {C.}~\bibnamefont {Lau}}, \bibinfo {author} {\bibfnamefont {B.}~\bibnamefont {Li}}, \bibinfo {author} {\bibfnamefont {S.~N.}\ \bibnamefont {Foo}}, \bibinfo {author} {\bibfnamefont {J.~O.}\ \bibnamefont {Yoo}}, \emph {et~al.},\ }in\ \href {https://doi.org/10.1364/OFC.2026.W1A.3} {\emph {\bibinfo {booktitle} {Optical {{Fiber Communication Conference}} ({{OFC}}) 2026}}}\ (\bibinfo  {publisher} {Optica Publishing Group},\ \bibinfo {address} {Los Angeles, California},\ \bibinfo {year} {2026})\ p.\ \bibinfo {pages} {W1A.3}\BibitemShut {NoStop}%
\bibitem [{\citenamefont {Wang}\ \emph {et~al.}(2024)\citenamefont {Wang}, \citenamefont {Li}, \citenamefont {Riemensberger}, \citenamefont {Lihachev}, \citenamefont {Churaev}, \citenamefont {Kao}, \citenamefont {Ji}, \citenamefont {Zhang}, \citenamefont {Blesin}, \citenamefont {Davydova}, \citenamefont {Chen}, \citenamefont {Huang}, \citenamefont {Wang}, \citenamefont {Ou},\ and\ \citenamefont {Kippenberg}}]{wang2024b}%
  \BibitemOpen
  \bibfield  {author} {\bibinfo {author} {\bibfnamefont {C.}~\bibnamefont {Wang}}, \bibinfo {author} {\bibfnamefont {Z.}~\bibnamefont {Li}}, \bibinfo {author} {\bibfnamefont {J.}~\bibnamefont {Riemensberger}}, \bibinfo {author} {\bibfnamefont {G.}~\bibnamefont {Lihachev}}, \bibinfo {author} {\bibfnamefont {M.}~\bibnamefont {Churaev}}, \emph {et~al.},\ }\href {https://doi.org/10.1038/s41586-024-07369-1} {\bibfield  {journal} {\bibinfo  {journal} {Nature}\ }\textbf {\bibinfo {volume} {629}},\ \bibinfo {pages} {784} (\bibinfo {year} {2024})}\BibitemShut {NoStop}%
\bibitem [{\citenamefont {Cai}\ \emph {et~al.}(2026)\citenamefont {Cai}, \citenamefont {Kotz}, \citenamefont {Larocque}, \citenamefont {Wang}, \citenamefont {Ji}, \citenamefont {Zhang}, \citenamefont {Drayss}, \citenamefont {Sun}, \citenamefont {Zheng}, \citenamefont {Ou}, \citenamefont {Koos},\ and\ \citenamefont {Kippenberg}}]{cai2026}%
  \BibitemOpen
  \bibfield  {author} {\bibinfo {author} {\bibfnamefont {J.}~\bibnamefont {Cai}}, \bibinfo {author} {\bibfnamefont {A.}~\bibnamefont {Kotz}}, \bibinfo {author} {\bibfnamefont {H.}~\bibnamefont {Larocque}}, \bibinfo {author} {\bibfnamefont {C.}~\bibnamefont {Wang}}, \bibinfo {author} {\bibfnamefont {X.}~\bibnamefont {Ji}}, \emph {et~al.},\ }\href {https://doi.org/10.1038/s41467-026-69769-3} {\bibfield  {journal} {\bibinfo  {journal} {Nature Communications}\ }\textbf {\bibinfo {volume} {17}},\ \bibinfo {pages} {3314} (\bibinfo {year} {2026})}\BibitemShut {NoStop}%
\bibitem [{\citenamefont {Adcock}\ \emph {et~al.}(2019)\citenamefont {Adcock}, \citenamefont {Vigliar}, \citenamefont {Santagati}, \citenamefont {Silverstone},\ and\ \citenamefont {Thompson}}]{Adcock2019a}%
  \BibitemOpen
  \bibfield  {author} {\bibinfo {author} {\bibfnamefont {J.~C.}\ \bibnamefont {Adcock}}, \bibinfo {author} {\bibfnamefont {C.}~\bibnamefont {Vigliar}}, \bibinfo {author} {\bibfnamefont {R.}~\bibnamefont {Santagati}}, \bibinfo {author} {\bibfnamefont {J.~W.}\ \bibnamefont {Silverstone}},\ and\ \bibinfo {author} {\bibfnamefont {M.~G.}\ \bibnamefont {Thompson}},\ }\bibfield  {journal} {\bibinfo  {journal} {Nature Communications}\ }\textbf {\bibinfo {volume} {10}},\ \href {https://doi.org/10.1038/s41467-019-11489-y} {10.1038/s41467-019-11489-y} (\bibinfo {year} {2019}),\ \Eprint {https://arxiv.org/abs/1811.03023} {arXiv:1811.03023} \BibitemShut {NoStop}%
\bibitem [{\citenamefont {Llewellyn}\ \emph {et~al.}(2020)\citenamefont {Llewellyn}, \citenamefont {Ding}, \citenamefont {Faruque}, \citenamefont {Paesani}, \citenamefont {Bacco}, \citenamefont {Santagati}, \citenamefont {Qian}, \citenamefont {Li}, \citenamefont {Xiao}, \citenamefont {Huber}, \citenamefont {Malik}, \citenamefont {Sinclair}, \citenamefont {Zhou}, \citenamefont {Rottwitt}, \citenamefont {O'Brien}, \citenamefont {Rarity}, \citenamefont {Gong}, \citenamefont {Oxenlowe}, \citenamefont {Wang},\ and\ \citenamefont {Thompson}}]{llewellyn2020}%
  \BibitemOpen
  \bibfield  {author} {\bibinfo {author} {\bibfnamefont {D.}~\bibnamefont {Llewellyn}}, \bibinfo {author} {\bibfnamefont {Y.}~\bibnamefont {Ding}}, \bibinfo {author} {\bibfnamefont {I.~I.}\ \bibnamefont {Faruque}}, \bibinfo {author} {\bibfnamefont {S.}~\bibnamefont {Paesani}}, \bibinfo {author} {\bibfnamefont {D.}~\bibnamefont {Bacco}}, \emph {et~al.},\ }\href {https://doi.org/10.1038/s41567-019-0727-x} {\bibfield  {journal} {\bibinfo  {journal} {Nature Physics}\ }\textbf {\bibinfo {volume} {16}},\ \bibinfo {pages} {148} (\bibinfo {year} {2020})},\ \Eprint {https://arxiv.org/abs/1911.07839} {arXiv:1911.07839} \BibitemShut {NoStop}%
\bibitem [{\citenamefont {Vigliar}\ \emph {et~al.}(2021)\citenamefont {Vigliar}, \citenamefont {Paesani}, \citenamefont {Ding}, \citenamefont {Adcock}, \citenamefont {Wang}, \citenamefont {{Morley-Short}}, \citenamefont {Bacco}, \citenamefont {Oxenl{\o}we}, \citenamefont {Thompson}, \citenamefont {Rarity},\ and\ \citenamefont {Laing}}]{vigliar2021}%
  \BibitemOpen
  \bibfield  {author} {\bibinfo {author} {\bibfnamefont {C.}~\bibnamefont {Vigliar}}, \bibinfo {author} {\bibfnamefont {S.}~\bibnamefont {Paesani}}, \bibinfo {author} {\bibfnamefont {Y.}~\bibnamefont {Ding}}, \bibinfo {author} {\bibfnamefont {J.~C.}\ \bibnamefont {Adcock}}, \bibinfo {author} {\bibfnamefont {J.}~\bibnamefont {Wang}}, \emph {et~al.},\ }\href {https://doi.org/10.1038/s41567-021-01333-w} {\bibfield  {journal} {\bibinfo  {journal} {Nature Physics}\ }\textbf {\bibinfo {volume} {17}},\ \bibinfo {pages} {1137} (\bibinfo {year} {2021})}\BibitemShut {NoStop}%
\bibitem [{\citenamefont {Bao}\ \emph {et~al.}(2023)\citenamefont {Bao}, \citenamefont {Fu}, \citenamefont {Pramanik}, \citenamefont {Mao}, \citenamefont {Chi}, \citenamefont {Cao}, \citenamefont {Zhai}, \citenamefont {Mao}, \citenamefont {Dai}, \citenamefont {Chen}, \citenamefont {Jia}, \citenamefont {Zhao}, \citenamefont {Zheng}, \citenamefont {Tang}, \citenamefont {Li}, \citenamefont {Luo}, \citenamefont {Wang}, \citenamefont {Yang}, \citenamefont {Peng}, \citenamefont {Liu}, \citenamefont {Dai}, \citenamefont {He}, \citenamefont {Muthali}, \citenamefont {Oxenl{\o}we}, \citenamefont {Vigliar}, \citenamefont {Paesani}, \citenamefont {Hou}, \citenamefont {Santagati}, \citenamefont {Silverstone}, \citenamefont {Laing}, \citenamefont {Thompson}, \citenamefont {O'Brien}, \citenamefont {Ding}, \citenamefont {Gong},\ and\ \citenamefont {Wang}}]{Bao2023}%
  \BibitemOpen
  \bibfield  {author} {\bibinfo {author} {\bibfnamefont {J.}~\bibnamefont {Bao}}, \bibinfo {author} {\bibfnamefont {Z.}~\bibnamefont {Fu}}, \bibinfo {author} {\bibfnamefont {T.}~\bibnamefont {Pramanik}}, \bibinfo {author} {\bibfnamefont {J.}~\bibnamefont {Mao}}, \bibinfo {author} {\bibfnamefont {Y.}~\bibnamefont {Chi}}, \emph {et~al.},\ }\href {https://doi.org/10.1038/s41566-023-01187-z} {\bibfield  {journal} {\bibinfo  {journal} {Nature Photonics}\ }\textbf {\bibinfo {volume} {17}},\ \bibinfo {pages} {573} (\bibinfo {year} {2023})}\BibitemShut {NoStop}%
\bibitem [{\citenamefont {Lee}\ \emph {et~al.}(2024)\citenamefont {Lee}, \citenamefont {Park}, \citenamefont {Bang}, \citenamefont {Sohn}, \citenamefont {Baldazzi}, \citenamefont {Sanna}, \citenamefont {Azzini},\ and\ \citenamefont {Pavesi}}]{Lee2024}%
  \BibitemOpen
  \bibfield  {author} {\bibinfo {author} {\bibfnamefont {J.-M.}\ \bibnamefont {Lee}}, \bibinfo {author} {\bibfnamefont {J.}~\bibnamefont {Park}}, \bibinfo {author} {\bibfnamefont {J.}~\bibnamefont {Bang}}, \bibinfo {author} {\bibfnamefont {Y.-I.}\ \bibnamefont {Sohn}}, \bibinfo {author} {\bibfnamefont {A.}~\bibnamefont {Baldazzi}}, \emph {et~al.},\ }\bibfield  {journal} {\bibinfo  {journal} {APL Photonics}\ }\textbf {\bibinfo {volume} {9}},\ \href {https://doi.org/10.1063/5.0207714} {10.1063/5.0207714} (\bibinfo {year} {2024})\BibitemShut {NoStop}%
\bibitem [{\citenamefont {Pont}\ \emph {et~al.}(2024)\citenamefont {Pont}, \citenamefont {Corrielli}, \citenamefont {Fyrillas}, \citenamefont {Agresti}, \citenamefont {Carvacho}, \citenamefont {Maring}, \citenamefont {Emeriau}, \citenamefont {Ceccarelli}, \citenamefont {Albiero}, \citenamefont {Dias~Ferreira}, \citenamefont {Somaschi}, \citenamefont {Senellart}, \citenamefont {Sagnes}, \citenamefont {Morassi}, \citenamefont {Lema{\^i}tre}, \citenamefont {Senellart}, \citenamefont {Sciarrino}, \citenamefont {Liscidini}, \citenamefont {Belabas},\ and\ \citenamefont {Osellame}}]{Pont2024}%
  \BibitemOpen
  \bibfield  {author} {\bibinfo {author} {\bibfnamefont {M.}~\bibnamefont {Pont}}, \bibinfo {author} {\bibfnamefont {G.}~\bibnamefont {Corrielli}}, \bibinfo {author} {\bibfnamefont {A.}~\bibnamefont {Fyrillas}}, \bibinfo {author} {\bibfnamefont {I.}~\bibnamefont {Agresti}}, \bibinfo {author} {\bibfnamefont {G.}~\bibnamefont {Carvacho}}, \emph {et~al.},\ }\href {https://doi.org/10.1038/s41534-024-00830-z} {\bibfield  {journal} {\bibinfo  {journal} {npj Quantum Information}\ }\textbf {\bibinfo {volume} {10}},\ \bibinfo {pages} {50} (\bibinfo {year} {2024})}\BibitemShut {NoStop}%
\bibitem [{\citenamefont {Huang}\ \emph {et~al.}(2024)\citenamefont {Huang}, \citenamefont {Li}, \citenamefont {Chen}, \citenamefont {Zhai}, \citenamefont {Zheng}, \citenamefont {Chi}, \citenamefont {Li}, \citenamefont {He}, \citenamefont {Gong},\ and\ \citenamefont {Wang}}]{huang2024}%
  \BibitemOpen
  \bibfield  {author} {\bibinfo {author} {\bibfnamefont {J.}~\bibnamefont {Huang}}, \bibinfo {author} {\bibfnamefont {X.}~\bibnamefont {Li}}, \bibinfo {author} {\bibfnamefont {X.}~\bibnamefont {Chen}}, \bibinfo {author} {\bibfnamefont {C.}~\bibnamefont {Zhai}}, \bibinfo {author} {\bibfnamefont {Y.}~\bibnamefont {Zheng}}, \emph {et~al.},\ }\href {https://doi.org/10.1038/s41467-024-46830-7} {\bibfield  {journal} {\bibinfo  {journal} {Nature Communications}\ }\textbf {\bibinfo {volume} {15}},\ \bibinfo {pages} {2601} (\bibinfo {year} {2024})}\BibitemShut {NoStop}%
\bibitem [{\citenamefont {Chen}\ \emph {et~al.}(2024{\natexlab{a}})\citenamefont {Chen}, \citenamefont {Wu}, \citenamefont {Lu}, \citenamefont {Wang}, \citenamefont {Lu}, \citenamefont {Zhu},\ and\ \citenamefont {Ma}}]{chen2024a}%
  \BibitemOpen
  \bibfield  {author} {\bibinfo {author} {\bibfnamefont {L.}~\bibnamefont {Chen}}, \bibinfo {author} {\bibfnamefont {B.}~\bibnamefont {Wu}}, \bibinfo {author} {\bibfnamefont {L.}~\bibnamefont {Lu}}, \bibinfo {author} {\bibfnamefont {K.}~\bibnamefont {Wang}}, \bibinfo {author} {\bibfnamefont {Y.}~\bibnamefont {Lu}}, \emph {et~al.},\ }\href {https://doi.org/10.1364/OE.515070} {\bibfield  {journal} {\bibinfo  {journal} {Optics Express}\ }\textbf {\bibinfo {volume} {32}},\ \bibinfo {pages} {14904} (\bibinfo {year} {2024}{\natexlab{a}})}\BibitemShut {NoStop}%
\bibitem [{\citenamefont {Chen}\ \emph {et~al.}(2024{\natexlab{b}})\citenamefont {Chen}, \citenamefont {Peng}, \citenamefont {Guo}, \citenamefont {Gu}, \citenamefont {Ding}, \citenamefont {Liu}, \citenamefont {Zhao}, \citenamefont {You}, \citenamefont {Qin}, \citenamefont {Wang}, \citenamefont {He}, \citenamefont {Renema}, \citenamefont {Huo}, \citenamefont {Wang}, \citenamefont {Lu},\ and\ \citenamefont {Pan}}]{chen2024}%
  \BibitemOpen
  \bibfield  {author} {\bibinfo {author} {\bibfnamefont {S.}~\bibnamefont {Chen}}, \bibinfo {author} {\bibfnamefont {L.-C.}\ \bibnamefont {Peng}}, \bibinfo {author} {\bibfnamefont {Y.-P.}\ \bibnamefont {Guo}}, \bibinfo {author} {\bibfnamefont {X.-M.}\ \bibnamefont {Gu}}, \bibinfo {author} {\bibfnamefont {X.}~\bibnamefont {Ding}}, \emph {et~al.},\ }\href {https://doi.org/10.1103/PhysRevLett.132.130603} {\bibfield  {journal} {\bibinfo  {journal} {Physical Review Letters}\ }\textbf {\bibinfo {volume} {132}},\ \bibinfo {pages} {130603} (\bibinfo {year} {2024}{\natexlab{b}})}\BibitemShut {NoStop}%
\bibitem [{\citenamefont {Maring}\ \emph {et~al.}(2024)\citenamefont {Maring}, \citenamefont {Fyrillas}, \citenamefont {Pont}, \citenamefont {Ivanov}, \citenamefont {Stepanov}, \citenamefont {Margaria}, \citenamefont {Hease}, \citenamefont {Pishchagin}, \citenamefont {Lema{\^i}tre}, \citenamefont {Sagnes}, \citenamefont {Au}, \citenamefont {Boissier}, \citenamefont {Bertasi}, \citenamefont {Baert}, \citenamefont {Valdivia}, \citenamefont {Billard}, \citenamefont {Acar}, \citenamefont {Brieussel}, \citenamefont {Mezher}, \citenamefont {Wein}, \citenamefont {Salavrakos}, \citenamefont {Sinnott}, \citenamefont {Fioretto}, \citenamefont {Emeriau}, \citenamefont {Belabas}, \citenamefont {Mansfield}, \citenamefont {Senellart}, \citenamefont {Senellart},\ and\ \citenamefont {Somaschi}}]{maring2024}%
  \BibitemOpen
  \bibfield  {author} {\bibinfo {author} {\bibfnamefont {N.}~\bibnamefont {Maring}}, \bibinfo {author} {\bibfnamefont {A.}~\bibnamefont {Fyrillas}}, \bibinfo {author} {\bibfnamefont {M.}~\bibnamefont {Pont}}, \bibinfo {author} {\bibfnamefont {E.}~\bibnamefont {Ivanov}}, \bibinfo {author} {\bibfnamefont {P.}~\bibnamefont {Stepanov}}, \emph {et~al.},\ }\href {https://doi.org/10.1038/s41566-024-01403-4} {\bibfield  {journal} {\bibinfo  {journal} {Nature Photonics}\ }\textbf {\bibinfo {volume} {18}},\ \bibinfo {pages} {603} (\bibinfo {year} {2024})}\BibitemShut {NoStop}%
\bibitem [{\citenamefont {Forbes}\ \emph {et~al.}(2025)\citenamefont {Forbes}, \citenamefont {Ghafari}, \citenamefont {Deacon}, \citenamefont {Singh}, \citenamefont {Lavie}, \citenamefont {Yard}, \citenamefont {Shaw}, \citenamefont {Laing},\ and\ \citenamefont {Tischler}}]{forbes2025}%
  \BibitemOpen
  \bibfield  {author} {\bibinfo {author} {\bibfnamefont {I.}~\bibnamefont {Forbes}}, \bibinfo {author} {\bibfnamefont {F.}~\bibnamefont {Ghafari}}, \bibinfo {author} {\bibfnamefont {E.~C.~R.}\ \bibnamefont {Deacon}}, \bibinfo {author} {\bibfnamefont {S.~P.}\ \bibnamefont {Singh}}, \bibinfo {author} {\bibfnamefont {E.}~\bibnamefont {Lavie}}, \emph {et~al.},\ }\href {https://doi.org/10.1088/1361-6633/adf85e} {\bibfield  {journal} {\bibinfo  {journal} {Reports on Progress in Physics}\ }\textbf {\bibinfo {volume} {88}},\ \bibinfo {pages} {086002} (\bibinfo {year} {2025})}\BibitemShut {NoStop}%
\bibitem [{\citenamefont {Carolan}\ \emph {et~al.}(2015)\citenamefont {Carolan}, \citenamefont {Harrold}, \citenamefont {Sparrow}, \citenamefont {{Mart{\'i}n-L{\'o}pez}}, \citenamefont {Russell}, \citenamefont {Silverstone}, \citenamefont {Shadbolt}, \citenamefont {Matsuda}, \citenamefont {Oguma}, \citenamefont {Itoh}, \citenamefont {Marshall}, \citenamefont {Thompson}, \citenamefont {Matthews}, \citenamefont {Hashimoto}, \citenamefont {O'Brien},\ and\ \citenamefont {Laing}}]{carolan2015}%
  \BibitemOpen
  \bibfield  {author} {\bibinfo {author} {\bibfnamefont {J.}~\bibnamefont {Carolan}}, \bibinfo {author} {\bibfnamefont {C.}~\bibnamefont {Harrold}}, \bibinfo {author} {\bibfnamefont {C.}~\bibnamefont {Sparrow}}, \bibinfo {author} {\bibfnamefont {E.}~\bibnamefont {{Mart{\'i}n-L{\'o}pez}}}, \bibinfo {author} {\bibfnamefont {N.~J.}\ \bibnamefont {Russell}}, \emph {et~al.},\ }\href {https://doi.org/10.1126/science.aab3642} {\bibfield  {journal} {\bibinfo  {journal} {Science}\ }\textbf {\bibinfo {volume} {349}},\ \bibinfo {pages} {711} (\bibinfo {year} {2015})}\BibitemShut {NoStop}%
\bibitem [{\citenamefont {Qiang}\ \emph {et~al.}(2018)\citenamefont {Qiang}, \citenamefont {Zhou}, \citenamefont {Wang}, \citenamefont {Wilkes}, \citenamefont {Loke}, \citenamefont {O'Gara}, \citenamefont {Kling}, \citenamefont {Marshall}, \citenamefont {Santagati}, \citenamefont {Ralph}, \citenamefont {Wang}, \citenamefont {O'Brien}, \citenamefont {Thompson},\ and\ \citenamefont {Matthews}}]{qiang2018}%
  \BibitemOpen
  \bibfield  {author} {\bibinfo {author} {\bibfnamefont {X.}~\bibnamefont {Qiang}}, \bibinfo {author} {\bibfnamefont {X.}~\bibnamefont {Zhou}}, \bibinfo {author} {\bibfnamefont {J.}~\bibnamefont {Wang}}, \bibinfo {author} {\bibfnamefont {C.~M.}\ \bibnamefont {Wilkes}}, \bibinfo {author} {\bibfnamefont {T.}~\bibnamefont {Loke}}, \emph {et~al.},\ }\href {https://doi.org/10.1038/s41566-018-0236-y} {\bibfield  {journal} {\bibinfo  {journal} {Nature Photonics}\ }\textbf {\bibinfo {volume} {12}},\ \bibinfo {pages} {534} (\bibinfo {year} {2018})}\BibitemShut {NoStop}%
\bibitem [{\citenamefont {Chi}\ \emph {et~al.}(2022)\citenamefont {Chi}, \citenamefont {Huang}, \citenamefont {Zhang}, \citenamefont {Mao}, \citenamefont {Zhou}, \citenamefont {Chen}, \citenamefont {Zhai}, \citenamefont {Bao}, \citenamefont {Dai}, \citenamefont {Yuan}, \citenamefont {Zhang}, \citenamefont {Dai}, \citenamefont {Tang}, \citenamefont {Yang}, \citenamefont {Li}, \citenamefont {Ding}, \citenamefont {Oxenl{\o}we}, \citenamefont {Thompson}, \citenamefont {O'Brien}, \citenamefont {Li}, \citenamefont {Gong},\ and\ \citenamefont {Wang}}]{chi2022}%
  \BibitemOpen
  \bibfield  {author} {\bibinfo {author} {\bibfnamefont {Y.}~\bibnamefont {Chi}}, \bibinfo {author} {\bibfnamefont {J.}~\bibnamefont {Huang}}, \bibinfo {author} {\bibfnamefont {Z.}~\bibnamefont {Zhang}}, \bibinfo {author} {\bibfnamefont {J.}~\bibnamefont {Mao}}, \bibinfo {author} {\bibfnamefont {Z.}~\bibnamefont {Zhou}}, \emph {et~al.},\ }\bibfield  {journal} {\bibinfo  {journal} {Nature Communications}\ }\textbf {\bibinfo {volume} {13}},\ \href {https://doi.org/10.1038/s41467-022-28767-x} {10.1038/s41467-022-28767-x} (\bibinfo {year} {2022})\BibitemShut {NoStop}%
\bibitem [{\citenamefont {Zhang}\ \emph {et~al.}(2023)\citenamefont {Zhang}, \citenamefont {Wan}, \citenamefont {Paesani}, \citenamefont {Laing}, \citenamefont {Shi}, \citenamefont {Cai}, \citenamefont {Luo}, \citenamefont {Lo}, \citenamefont {Kwek},\ and\ \citenamefont {Liu}}]{zhang2023}%
  \BibitemOpen
  \bibfield  {author} {\bibinfo {author} {\bibfnamefont {H.}~\bibnamefont {Zhang}}, \bibinfo {author} {\bibfnamefont {L.}~\bibnamefont {Wan}}, \bibinfo {author} {\bibfnamefont {S.}~\bibnamefont {Paesani}}, \bibinfo {author} {\bibfnamefont {A.}~\bibnamefont {Laing}}, \bibinfo {author} {\bibfnamefont {Y.}~\bibnamefont {Shi}}, \emph {et~al.},\ }\href {https://doi.org/10.1103/PRXQuantum.4.030340} {\bibfield  {journal} {\bibinfo  {journal} {PRX Quantum}\ }\textbf {\bibinfo {volume} {4}},\ \bibinfo {pages} {030340} (\bibinfo {year} {2023})}\BibitemShut {NoStop}%
\bibitem [{\citenamefont {Ciampini}\ \emph {et~al.}(2016)\citenamefont {Ciampini}, \citenamefont {Orieux}, \citenamefont {Paesani}, \citenamefont {Sciarrino}, \citenamefont {Corrielli}, \citenamefont {Crespi}, \citenamefont {Ramponi}, \citenamefont {Osellame},\ and\ \citenamefont {Mataloni}}]{ciampini2016}%
  \BibitemOpen
  \bibfield  {author} {\bibinfo {author} {\bibfnamefont {M.~A.}\ \bibnamefont {Ciampini}}, \bibinfo {author} {\bibfnamefont {A.}~\bibnamefont {Orieux}}, \bibinfo {author} {\bibfnamefont {S.}~\bibnamefont {Paesani}}, \bibinfo {author} {\bibfnamefont {F.}~\bibnamefont {Sciarrino}}, \bibinfo {author} {\bibfnamefont {G.}~\bibnamefont {Corrielli}}, \emph {et~al.},\ }\href {https://doi.org/10.1038/lsa.2016.64} {\bibfield  {journal} {\bibinfo  {journal} {Light: Science \& Applications}\ }\textbf {\bibinfo {volume} {5}},\ \bibinfo {pages} {e16064} (\bibinfo {year} {2016})}\BibitemShut {NoStop}%
\bibitem [{\citenamefont {Aghaee~Rad}\ \emph {et~al.}(2025)\citenamefont {Aghaee~Rad}, \citenamefont {Ainsworth}, \citenamefont {Alexander}, \citenamefont {Altieri}, \citenamefont {Askarani}, \citenamefont {Baby}, \citenamefont {Banchi}, \citenamefont {Baragiola}, \citenamefont {Bourassa}, \citenamefont {Chadwick}, \citenamefont {Charania}, \citenamefont {Chen}, \citenamefont {Collins}, \citenamefont {Contu}, \citenamefont {D'Arcy}, \citenamefont {Dauphinais}, \citenamefont {De~Prins}, \citenamefont {Deschenes}, \citenamefont {Di~Luch}, \citenamefont {Duque}, \citenamefont {Edke}, \citenamefont {Fayer}, \citenamefont {Ferracin}, \citenamefont {Ferretti}, \citenamefont {Gefaell}, \citenamefont {Glancy}, \citenamefont {{Gonz{\'a}lez-Arciniegas}}, \citenamefont {Grainge}, \citenamefont {Han}, \citenamefont {Hastrup}, \citenamefont {Helt}, \citenamefont {Hillmann}, \citenamefont {Hundal}, \citenamefont {Izumi}, \citenamefont {Jaeken}, \citenamefont {Jonas}, \citenamefont {Kocsis}, \citenamefont {Krasnokutska}, \citenamefont {Larsen}, \citenamefont {Laskowski}, \citenamefont {Laudenbach}, \citenamefont {Lavoie}, \citenamefont {Li}, \citenamefont {Lomonte}, \citenamefont {Lopetegui}, \citenamefont {Luey}, \citenamefont {Lund}, \citenamefont {Ma}, \citenamefont {Madsen}, \citenamefont {Mahler}, \citenamefont {Mantilla~Calder{\'o}n}, \citenamefont {Menotti}, \citenamefont {Miatto}, \citenamefont {Morrison}, \citenamefont {Nadkarni}, \citenamefont {Nakamura}, \citenamefont {Neuhaus}, \citenamefont {Niu}, \citenamefont {Noro}, \citenamefont {Papirov}, \citenamefont {Pesah}, \citenamefont {Phillips}, \citenamefont {Plick}, \citenamefont {Rogalsky}, \citenamefont {Rortais}, \citenamefont {{Sabines-Chesterking}}, \citenamefont {{Safavi-Bayat}}, \citenamefont {Sazhaev}, \citenamefont {Seymour}, \citenamefont {Rezaei~Shad}, \citenamefont {Silverman}, \citenamefont {Srinivasan}, \citenamefont {Stephan}, \citenamefont {Tang}, \citenamefont {Tasker}, \citenamefont {Teo}, \citenamefont {Then}, \citenamefont {Tremblay}, \citenamefont {Tzitrin}, \citenamefont {Vaidya}, \citenamefont {Vasmer}, \citenamefont {Vernon}, \citenamefont {Villalobos}, \citenamefont {Walshe}, \citenamefont {Weil}, \citenamefont {Xin}, \citenamefont {Yan}, \citenamefont {Yao}, \citenamefont {Zamani~Abnili},\ and\ \citenamefont {Zhang}}]{aghaeerad2025}%
  \BibitemOpen
  \bibfield  {author} {\bibinfo {author} {\bibfnamefont {H.}~\bibnamefont {Aghaee~Rad}}, \bibinfo {author} {\bibfnamefont {T.}~\bibnamefont {Ainsworth}}, \bibinfo {author} {\bibfnamefont {R.~N.}\ \bibnamefont {Alexander}}, \bibinfo {author} {\bibfnamefont {B.}~\bibnamefont {Altieri}}, \bibinfo {author} {\bibfnamefont {M.~F.}\ \bibnamefont {Askarani}}, \emph {et~al.},\ }\href {https://doi.org/10.1038/s41586-024-08406-9} {\bibfield  {journal} {\bibinfo  {journal} {Nature}\ }\textbf {\bibinfo {volume} {638}},\ \bibinfo {pages} {912} (\bibinfo {year} {2025})}\BibitemShut {NoStop}%
\bibitem [{\citenamefont {Jia}\ \emph {et~al.}(2025)\citenamefont {Jia}, \citenamefont {Zhai}, \citenamefont {Zhu}, \citenamefont {You}, \citenamefont {Cao}, \citenamefont {Zhang}, \citenamefont {Zheng}, \citenamefont {Fu}, \citenamefont {Mao}, \citenamefont {Dai}, \citenamefont {Chang}, \citenamefont {Su}, \citenamefont {Gong},\ and\ \citenamefont {Wang}}]{jia2025}%
  \BibitemOpen
  \bibfield  {author} {\bibinfo {author} {\bibfnamefont {X.}~\bibnamefont {Jia}}, \bibinfo {author} {\bibfnamefont {C.}~\bibnamefont {Zhai}}, \bibinfo {author} {\bibfnamefont {X.}~\bibnamefont {Zhu}}, \bibinfo {author} {\bibfnamefont {C.}~\bibnamefont {You}}, \bibinfo {author} {\bibfnamefont {Y.}~\bibnamefont {Cao}}, \emph {et~al.},\ }\href {https://doi.org/10.1038/s41586-025-08602-1} {\bibfield  {journal} {\bibinfo  {journal} {Nature}\ }\textbf {\bibinfo {volume} {639}},\ \bibinfo {pages} {329} (\bibinfo {year} {2025})}\BibitemShut {NoStop}%
\bibitem [{\citenamefont {Wang}\ \emph {et~al.}(2025)\citenamefont {Wang}, \citenamefont {Li}, \citenamefont {Wang}, \citenamefont {Zhou}, \citenamefont {Cheng}, \citenamefont {Jing}, \citenamefont {Sun}, \citenamefont {Li}, \citenamefont {Li}, \citenamefont {Wu}, \citenamefont {Gong}, \citenamefont {He}, \citenamefont {Li},\ and\ \citenamefont {Yang}}]{wang2025a}%
  \BibitemOpen
  \bibfield  {author} {\bibinfo {author} {\bibfnamefont {Z.}~\bibnamefont {Wang}}, \bibinfo {author} {\bibfnamefont {K.}~\bibnamefont {Li}}, \bibinfo {author} {\bibfnamefont {Y.}~\bibnamefont {Wang}}, \bibinfo {author} {\bibfnamefont {X.}~\bibnamefont {Zhou}}, \bibinfo {author} {\bibfnamefont {Y.}~\bibnamefont {Cheng}}, \emph {et~al.},\ }\href {https://doi.org/10.1038/s41377-025-01812-2} {\bibfield  {journal} {\bibinfo  {journal} {Light: Science \& Applications}\ }\textbf {\bibinfo {volume} {14}},\ \bibinfo {pages} {164} (\bibinfo {year} {2025})}\BibitemShut {NoStop}%
\bibitem [{\citenamefont {Jia}\ \emph {et~al.}(2026)\citenamefont {Jia}, \citenamefont {You}, \citenamefont {Zhai}, \citenamefont {Zhu}, \citenamefont {Zheng}, \citenamefont {Dai}, \citenamefont {Fu}, \citenamefont {Su}, \citenamefont {Gong},\ and\ \citenamefont {Wang}}]{jia2026}%
  \BibitemOpen
  \bibfield  {author} {\bibinfo {author} {\bibfnamefont {X.}~\bibnamefont {Jia}}, \bibinfo {author} {\bibfnamefont {C.}~\bibnamefont {You}}, \bibinfo {author} {\bibfnamefont {C.}~\bibnamefont {Zhai}}, \bibinfo {author} {\bibfnamefont {X.}~\bibnamefont {Zhu}}, \bibinfo {author} {\bibfnamefont {Y.}~\bibnamefont {Zheng}}, \emph {et~al.},\ }\href {https://doi.org/10.1038/s41566-026-01868-5} {\bibfield  {journal} {\bibinfo  {journal} {Nature Photonics}\ }\textbf {\bibinfo {volume} {20}},\ \bibinfo {pages} {428} (\bibinfo {year} {2026})}\BibitemShut {NoStop}%
\bibitem [{\citenamefont {Grover}(1996)}]{grover1996}%
  \BibitemOpen
  \bibfield  {author} {\bibinfo {author} {\bibfnamefont {L.~K.}\ \bibnamefont {Grover}},\ }in\ \href {https://doi.org/10.1145/237814.237866} {\emph {\bibinfo {booktitle} {Proceedings of the Twenty-Eighth Annual {{ACM}} Symposium on {{Theory}} of Computing - {{STOC}} '96}}}\ (\bibinfo  {publisher} {ACM Press},\ \bibinfo {address} {New York, New York, USA},\ \bibinfo {year} {1996})\ pp.\ \bibinfo {pages} {212--219}\BibitemShut {NoStop}%
\bibitem [{\citenamefont {Grover}(1997)}]{grover1997}%
  \BibitemOpen
  \bibfield  {author} {\bibinfo {author} {\bibfnamefont {L.~K.}\ \bibnamefont {Grover}},\ }\href {https://doi.org/10.1103/PhysRevLett.79.325} {\bibfield  {journal} {\bibinfo  {journal} {Physical Review Letters}\ }\textbf {\bibinfo {volume} {79}},\ \bibinfo {pages} {325} (\bibinfo {year} {1997})}\BibitemShut {NoStop}%
\bibitem [{\citenamefont {Deutsch}\ and\ \citenamefont {Jozsa}(1992)}]{deutsch1992}%
  \BibitemOpen
  \bibfield  {author} {\bibinfo {author} {\bibfnamefont {D.}~\bibnamefont {Deutsch}}\ and\ \bibinfo {author} {\bibfnamefont {R.}~\bibnamefont {Jozsa}},\ }\href {https://doi.org/10.1098/rspa.1992.0167} {\bibfield  {journal} {\bibinfo  {journal} {Proceedings of the Royal Society of London. Series A: Mathematical and Physical Sciences}\ }\textbf {\bibinfo {volume} {439}},\ \bibinfo {pages} {553} (\bibinfo {year} {1992})}\BibitemShut {NoStop}%
\bibitem [{\citenamefont {Dhand}\ \emph {et~al.}(2018)\citenamefont {Dhand}, \citenamefont {Engelkemeier}, \citenamefont {Sansoni}, \citenamefont {Barkhofen}, \citenamefont {Silberhorn},\ and\ \citenamefont {Plenio}}]{dhand2018}%
  \BibitemOpen
  \bibfield  {author} {\bibinfo {author} {\bibfnamefont {I.}~\bibnamefont {Dhand}}, \bibinfo {author} {\bibfnamefont {M.}~\bibnamefont {Engelkemeier}}, \bibinfo {author} {\bibfnamefont {L.}~\bibnamefont {Sansoni}}, \bibinfo {author} {\bibfnamefont {S.}~\bibnamefont {Barkhofen}}, \bibinfo {author} {\bibfnamefont {C.}~\bibnamefont {Silberhorn}}, \emph {et~al.},\ }\href {https://doi.org/10.1103/PhysRevLett.120.130501} {\bibfield  {journal} {\bibinfo  {journal} {Physical Review Letters}\ }\textbf {\bibinfo {volume} {120}},\ \bibinfo {pages} {130501} (\bibinfo {year} {2018})}\BibitemShut {NoStop}%
\bibitem [{\citenamefont {{Meyer-Scott}}\ \emph {et~al.}(2022)\citenamefont {{Meyer-Scott}}, \citenamefont {Prasannan}, \citenamefont {Dhand}, \citenamefont {Eigner}, \citenamefont {Quiring}, \citenamefont {Barkhofen}, \citenamefont {Brecht}, \citenamefont {Plenio},\ and\ \citenamefont {Silberhorn}}]{meyer-scott2022}%
  \BibitemOpen
  \bibfield  {author} {\bibinfo {author} {\bibfnamefont {E.}~\bibnamefont {{Meyer-Scott}}}, \bibinfo {author} {\bibfnamefont {N.}~\bibnamefont {Prasannan}}, \bibinfo {author} {\bibfnamefont {I.}~\bibnamefont {Dhand}}, \bibinfo {author} {\bibfnamefont {C.}~\bibnamefont {Eigner}}, \bibinfo {author} {\bibfnamefont {V.}~\bibnamefont {Quiring}}, \emph {et~al.},\ }\href {https://doi.org/10.1103/PhysRevLett.129.150501} {\bibfield  {journal} {\bibinfo  {journal} {Physical Review Letters}\ }\textbf {\bibinfo {volume} {129}},\ \bibinfo {pages} {150501} (\bibinfo {year} {2022})}\BibitemShut {NoStop}%
\bibitem [{\citenamefont {G{\"u}hne}\ and\ \citenamefont {T{\'o}th}(2009)}]{guhne2009}%
  \BibitemOpen
  \bibfield  {author} {\bibinfo {author} {\bibfnamefont {O.}~\bibnamefont {G{\"u}hne}}\ and\ \bibinfo {author} {\bibfnamefont {G.}~\bibnamefont {T{\'o}th}},\ }\href {https://doi.org/10.1016/j.physrep.2009.02.004} {\bibfield  {journal} {\bibinfo  {journal} {Physics Reports}\ }\textbf {\bibinfo {volume} {474}},\ \bibinfo {pages} {1} (\bibinfo {year} {2009})},\ \Eprint {https://arxiv.org/abs/0811.2803} {arXiv:0811.2803} \BibitemShut {NoStop}%
\bibitem [{\citenamefont {T{\'o}th}\ and\ \citenamefont {G{\"u}hne}(2005)}]{toth2005}%
  \BibitemOpen
  \bibfield  {author} {\bibinfo {author} {\bibfnamefont {G.}~\bibnamefont {T{\'o}th}}\ and\ \bibinfo {author} {\bibfnamefont {O.}~\bibnamefont {G{\"u}hne}},\ }\href {https://doi.org/10.1103/PhysRevLett.94.060501} {\bibfield  {journal} {\bibinfo  {journal} {Physical Review Letters}\ }\textbf {\bibinfo {volume} {94}},\ \bibinfo {pages} {060501} (\bibinfo {year} {2005})}\BibitemShut {NoStop}%
\bibitem [{\citenamefont {Cabello}\ \emph {et~al.}(2008)\citenamefont {Cabello}, \citenamefont {G{\"u}hne},\ and\ \citenamefont {Rodr{\'i}guez}}]{cabello2008}%
  \BibitemOpen
  \bibfield  {author} {\bibinfo {author} {\bibfnamefont {A.}~\bibnamefont {Cabello}}, \bibinfo {author} {\bibfnamefont {O.}~\bibnamefont {G{\"u}hne}},\ and\ \bibinfo {author} {\bibfnamefont {D.}~\bibnamefont {Rodr{\'i}guez}},\ }\href {https://doi.org/10.1103/PhysRevA.77.062106} {\bibfield  {journal} {\bibinfo  {journal} {Physical Review A}\ }\textbf {\bibinfo {volume} {77}},\ \bibinfo {pages} {062106} (\bibinfo {year} {2008})}\BibitemShut {NoStop}%
\bibitem [{\citenamefont {Kim}\ \emph {et~al.}(2025)\citenamefont {Kim}, \citenamefont {Kaupp}, \citenamefont {Reum}, \citenamefont {Peniakov}, \citenamefont {Michl}, \citenamefont {Kohr}, \citenamefont {Emmerling}, \citenamefont {Kamp}, \citenamefont {Cho}, \citenamefont {Huber-Loyola}, \citenamefont {H{\"o}fling},\ and\ \citenamefont {Pfenning}}]{Kim2025}%
  \BibitemOpen
  \bibfield  {author} {\bibinfo {author} {\bibfnamefont {J.}~\bibnamefont {Kim}}, \bibinfo {author} {\bibfnamefont {J.}~\bibnamefont {Kaupp}}, \bibinfo {author} {\bibfnamefont {Y.}~\bibnamefont {Reum}}, \bibinfo {author} {\bibfnamefont {G.}~\bibnamefont {Peniakov}}, \bibinfo {author} {\bibfnamefont {J.}~\bibnamefont {Michl}}, \emph {et~al.},\ }\bibfield  {journal} {\bibinfo  {journal} {Advanced Quantum Technologies}\ }\textbf {\bibinfo {volume} {8}},\ \href {https://doi.org/10.1002/qute.202500069} {10.1002/qute.202500069} (\bibinfo {year} {2025})\BibitemShut {NoStop}%
\bibitem [{\citenamefont {Pfister}\ \emph {et~al.}(2025)\citenamefont {Pfister}, \citenamefont {Wendland}, \citenamefont {Hornung}, \citenamefont {Engel}, \citenamefont {H{\"u}ging}, \citenamefont {Herzog}, \citenamefont {Vijayan}, \citenamefont {Joos}, \citenamefont {Jung}, \citenamefont {Jetter}, \citenamefont {Portalupi}, \citenamefont {Pernice},\ and\ \citenamefont {Michler}}]{Pfister2025}%
  \BibitemOpen
  \bibfield  {author} {\bibinfo {author} {\bibfnamefont {U.}~\bibnamefont {Pfister}}, \bibinfo {author} {\bibfnamefont {D.}~\bibnamefont {Wendland}}, \bibinfo {author} {\bibfnamefont {F.}~\bibnamefont {Hornung}}, \bibinfo {author} {\bibfnamefont {L.}~\bibnamefont {Engel}}, \bibinfo {author} {\bibfnamefont {H.}~\bibnamefont {H{\"u}ging}}, \emph {et~al.},\ }\href {https://doi.org/10.1038/s44310-025-00061-w} {\bibfield  {journal} {\bibinfo  {journal} {npj Nanophotonics}\ }\textbf {\bibinfo {volume} {2}},\ \bibinfo {pages} {11} (\bibinfo {year} {2025})}\BibitemShut {NoStop}%
\bibitem [{\citenamefont {Hauser}\ \emph {et~al.}(2026)\citenamefont {Hauser}, \citenamefont {Bayerbach}, \citenamefont {Kaupp}, \citenamefont {Reum}, \citenamefont {Peniakov}, \citenamefont {Michl}, \citenamefont {Kamp}, \citenamefont {{Huber-Loyola}}, \citenamefont {Pfenning}, \citenamefont {H{\"o}fling},\ and\ \citenamefont {Barz}}]{Hauser2026}%
  \BibitemOpen
  \bibfield  {author} {\bibinfo {author} {\bibfnamefont {N.}~\bibnamefont {Hauser}}, \bibinfo {author} {\bibfnamefont {M.}~\bibnamefont {Bayerbach}}, \bibinfo {author} {\bibfnamefont {J.}~\bibnamefont {Kaupp}}, \bibinfo {author} {\bibfnamefont {Y.}~\bibnamefont {Reum}}, \bibinfo {author} {\bibfnamefont {G.}~\bibnamefont {Peniakov}}, \emph {et~al.},\ }\href {https://doi.org/10.1038/s41467-026-68336-0} {\bibfield  {journal} {\bibinfo  {journal} {Nature Communications}\ }\textbf {\bibinfo {volume} {17}},\ \bibinfo {pages} {537} (\bibinfo {year} {2026})}\BibitemShut {NoStop}%
\bibitem [{\citenamefont {Bartolucci}\ \emph {et~al.}(2021)\citenamefont {Bartolucci}, \citenamefont {Birchall}, \citenamefont {Bonneau}, \citenamefont {Cable}, \citenamefont {{Gimeno-Segovia}}, \citenamefont {Kieling}, \citenamefont {Nickerson}, \citenamefont {Rudolph},\ and\ \citenamefont {Sparrow}}]{PsiQSwitch2021}%
  \BibitemOpen
  \bibfield  {author} {\bibinfo {author} {\bibfnamefont {S.}~\bibnamefont {Bartolucci}}, \bibinfo {author} {\bibfnamefont {P.}~\bibnamefont {Birchall}}, \bibinfo {author} {\bibfnamefont {D.}~\bibnamefont {Bonneau}}, \bibinfo {author} {\bibfnamefont {H.}~\bibnamefont {Cable}}, \bibinfo {author} {\bibfnamefont {M.}~\bibnamefont {{Gimeno-Segovia}}}, \emph {et~al.},\ }\href@noop {} {\bibinfo {title} {Switch networks for photonic fusion-based quantum computing}} (\bibinfo {year} {2021}),\ \Eprint {https://arxiv.org/abs/2109.13760} {arXiv:2109.13760} \BibitemShut {NoStop}%
\bibitem [{\citenamefont {Cao}\ \emph {et~al.}(2024)\citenamefont {Cao}, \citenamefont {Hansen}, \citenamefont {Giorgino}, \citenamefont {Carosini}, \citenamefont {Zah{\'a}lka}, \citenamefont {Zilk}, \citenamefont {Loredo},\ and\ \citenamefont {Walther}}]{Cao2024}%
  \BibitemOpen
  \bibfield  {author} {\bibinfo {author} {\bibfnamefont {H.}~\bibnamefont {Cao}}, \bibinfo {author} {\bibfnamefont {L.~M.}\ \bibnamefont {Hansen}}, \bibinfo {author} {\bibfnamefont {F.}~\bibnamefont {Giorgino}}, \bibinfo {author} {\bibfnamefont {L.}~\bibnamefont {Carosini}}, \bibinfo {author} {\bibfnamefont {P.}~\bibnamefont {Zah{\'a}lka}}, \emph {et~al.},\ }\href {https://doi.org/10.1103/PhysRevLett.132.130604} {\bibfield  {journal} {\bibinfo  {journal} {Physical Review Letters}\ }\textbf {\bibinfo {volume} {132}},\ \bibinfo {pages} {130604} (\bibinfo {year} {2024})}\BibitemShut {NoStop}%
\bibitem [{\citenamefont {Beck}(2007)}]{Beck2007}%
  \BibitemOpen
  \bibfield  {author} {\bibinfo {author} {\bibfnamefont {M.}~\bibnamefont {Beck}},\ }\href {https://doi.org/10.1364/JOSAB.24.002972} {\bibfield  {journal} {\bibinfo  {journal} {Journal of the Optical Society of America B}\ }\textbf {\bibinfo {volume} {24}},\ \bibinfo {pages} {2972} (\bibinfo {year} {2007})}\BibitemShut {NoStop}%
\end{thebibliography}%
\newpage
\section{Appendix}
    
    \subsection{The photonic circuit}
        The chip utilises $450$\texttimes $\qty{220}{\nano\metre}$ single-mode waveguides, grating couplers with backside mirrors for higher coupling efficiencies, and titanium-nitride thermo-optic phase shifters with deep trenches, to mitigate thermal crosstalk.\\ 
        We estimate the waveguide transmission via the cut-back method and obtain 
        $\eta_{WG}= \qty{-6.5+-0.3}{\decibel\per\centi\metre}$. 
        One contributing factor to this loss is periodicity of the side wall roughness stemming from electron-beam writing patterns.
        The grating couplers reach a \gls{ce} of \qty{-1.2+-0.2}{\decibel} $\approx$ \qty{75+-4}{\percent} at 
        \qty{1550}{\nano\metre} (excluding waveguide losses).
        From these values, and assuming $\qty{0.2}{\decibel}$ loss per \gls{mmi}, we expect an overall transmission per mode of $\qty{-7.3+-0.4}{\decibel} \approx \qty{18+-2}{\percent}$.
        In comparison, the recorded four-fold single-photon data suggests an average transmission per mode of $\qty{-8.9}{\decibel} \approx \qty{12.8}{\percent}$. 
        This discrepancy can be attributed to time varying \glspl{ce}, caused by fibre array alignment drifts,
        which we mitigate through automated realignment routines at fixed time intervals. \\

    \subsection{Single-photon source}
         
        In the pump power regime used for the experiments, our single-photon source exhibits a heralded 
        second-order correlation of $g^{(2)}(0)=\qty{0.030\pm0.002}{}$ \cite{Beck2007}, which translates into a pair-generation amplitude of $\lambda = \text{tanh}(r) = \qty{0.085\pm0.003}{}$, where $r$ is typically referred to as the squeezing parameter.
        At this pump power, and filtered by $\qty{1.5}{\nano\metre}$ band pass filters, we obtain single-source visibilities of \qty{98.4 \pm 0.1}{\percent} and a heralded two-source \gls{hom} visibility of \qty{93.7 \pm 1.3}{\percent}. 
        The non-perfect single-source visibility is mainly due to higher order emissions in combination with loss, and additional residual spectral correlations result in the lower two-source \gls{hom} visibility.     
    
    \subsection{Graph state generation and verification}\label{sec:state_gen}
        The two four-qubit states created by our photonic integrated circuit are 
        a \gls{ghz} state $\state{\text{GHZ}_4}$ and $\state{\text{L}_4}$, which are locally equivalent to a four-qubit star $\state{\text{Star}_4}$ and a linear graph state $\state{\text{Lin}_4}$:
        \begin{align}
            \state{\text{L}_4} &= H_1 Z_1 Z_2 Z_3 \state{\text{Lin}_4} \notag \\ 
                          &= H_1 Z_1 Z_2 Z_3 CZ_{12} \text{C}Z_{23} \text{C}Z_{34} \state{\p\p\p\p}\\      
            \state{\text{GHZ}_4} &=\frac{1}{\sqrt{2}} \left(\state{0101} + \state{1010}\right)  \\
                          &= H_2 Z_2 H_3 H_4 Z_4 \state{\text{Star}_4} \notag \\
                          &= H_2 Z_2 H_3 H_4 Z_4 \text{C}Z_{12} \text{C}Z_{13} \text{C}Z_{14} \state{\p\p\p\p} \notag
        \end{align}
        The box graph state $\state{\square_4}$ utilised for the Grover's search algorithm is related to the state $\state{\text{L}_4}$ as
        \begin{align}
            \state{\square_4} &= \text{SWAP}_{23}(HX)_1 (HZ)_2 (HZ)_3\state{\text{L}_4},
        \end{align}
        where the swap gate is realised by swapping the information from qubit 2 and qubit 3 in post-processing.\\
        An $n$-qubit graph state $\state{G}$ has $2^n$ stabilisers.
        Measuring all 16 stabiliser expectation values for the generated graph states, requires only 9 measurement settings. 
        The measurement settings are listed in Tab.~\ref{tab:stabilisers}.\\  
         \begin{table}[h!]
            \centering
            \caption{Measurement bases for stabiliser expectation values.
                     \label{tab:stabilisers}}
            \begin{tabular}{|c|c|c|}
            \hline
            State & Basis & Stabiliser expectation values\\
            \hline
            $\state{{GHZ}_4}$ & $ZZZZ$     & $\exval{-11ZZ}, \exval{1Z1Z}, \exval{-1ZZ1},
                                             \exval{-Z11Z}$,\\
                              &            & $\exval{Z1Z1}, \exval{-ZZ11}, \exval{ZZZZ}$ \\
                              & $XXXX$     & $\exval{XXXX}$\\
                              & $XXYY$     & $\exval{XXYY}$\\
                              & $XYXY$     & $\exval{-XYXY}$\\
                              & $XYYX$     & $\exval{XYYX}$\\ 
                              & $YXXY$     & $\exval{YXXY}$\\
                              & $YXYX$     & $\exval{-YXYX}$\\
                              & $YYXX$     & $\exval{YYXX}$\\
                              & $YYYY$     & $\exval{YYYY}$\\
                              \hline
            $\state{{L}_4}$   & $XXZX$     & $\exval{11ZX}, \exval{-XX1X}, \exval{-XXZ1}$\\
                              & $YYZX$     & $\exval{-YY1X}, \exval{-YYZ1}$\\
                              & $ZZXZ$     & $\exval{-1ZXZ}, \exval{Z1XZ}$\\
                              & $ZZYY$     & $\exval{-1ZYY}, \exval{Z1YY}$\\
                              & $ZZZX$     & $\exval{-ZZ11}, \exval{-ZZZX}$\\ 
                              & $XYXY$     & $\exval{-XYXY}$\\
                              & $XYYZ$     & $\exval{XYYZ}$\\
                              & $YXXY$     & $\exval{YXXY}$\\
                              & $YXYZ$     & $\exval{-YXYZ}$\\
            \hline
            \end{tabular}
        \end{table}    
            \begin{figure*}[htbp]
            \includegraphics[width=\textwidth]{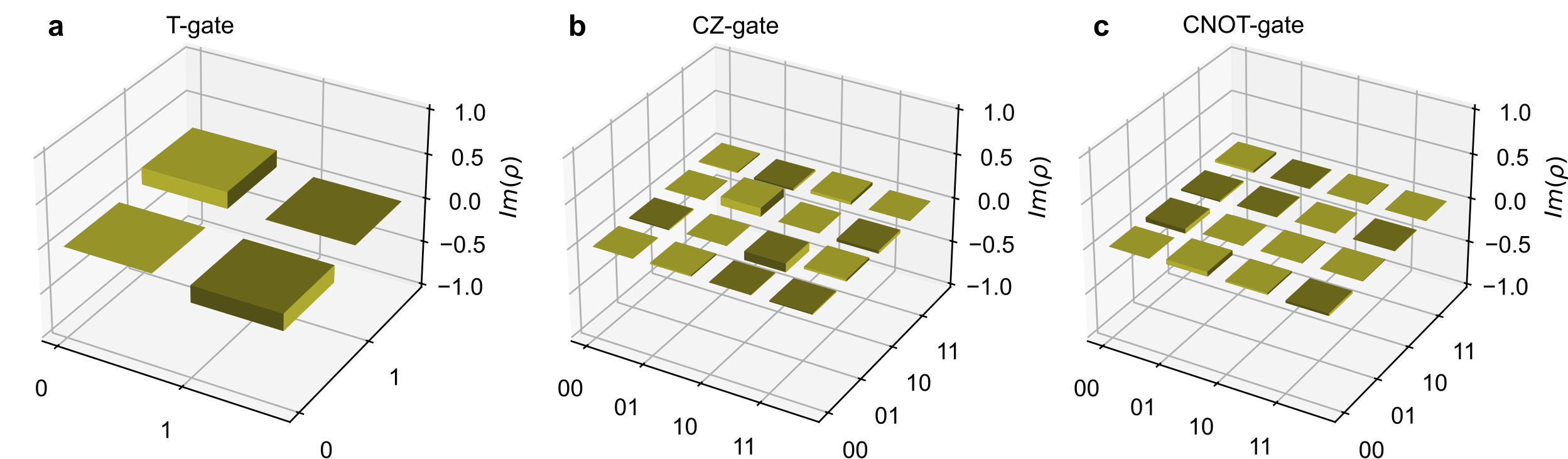}
            \caption{Imaginary parts of the density matrices $\rho$ for the output states of the \gls{mbqc} implemented single-qubit and two-qubit gates (real parts shown in Fig. 3 main text).
            \label{fig:mbqc_gates_im}} 
        \end{figure*} 
        In the main text the stabiliser were abbreviated as $s_i$, where for the state $\state{\text{GHZ}_4}$:
        $s_{ 0 }= 1111$, $s_{ 1 }= -11ZZ$, $s_{ 2 }= 1Z1Z$, $s_{ 3 }= -1ZZ1$, $s_{ 4 }= XXXX$, $s_{ 5 }= XXYY$, $s_{ 6 }= -XYXY$, $s_{ 7 }= XYYX$, $s_{ 8 }=YXXY$, $s_{ 9 }= -YXYX$, $s_{ 10 }= YYXX$, $s_{ 11 }= YYYY$, $s_{ 12 }= -Z11Z$, $s_{ 13 }= Z1Z1$, $s_{ 14 }= -ZZ11$, $s_{ 15 }= ZZZZ$.\\
        and for the state $\state{\text{L}_{4}}$: 
        $s_{ 0 }= 1111$, $s_{ 1 }= 11ZX$, $s_{ 2 }= -1ZXZ$, $s_{ 3 }= -1ZYY$, $s_{ 4 }= -XX1X$, $s_{ 5 }= -XXZ1$, $s_{ 6 }= -XYXY$, $s_{ 7 }= XYYZ$, $s_{ 8 }= YXXY$, $s_{ 9 }= -YXYZ$, $s_{ 10 }= -YY1X$, $s_{ 11 }= -YYZ1$, $s_{ 12 }= Z1XZ$, $s_{ 13 }= Z1YY$, $s_{ 14 }=-ZZ11$, $s_{ 15 }= -ZZZX$.

        Uncertainties for the stabiliser expectation values were calculated using Monte Carlo simulations.
        We assumed Possionian-distributed counts and generated 1000 simulations per data set.

    \subsection{Bell inequalities}
        The two-setting Bell inequality $|\exval{\inequal_{\text{II}}}|\leq 2$ 
        \cite{cabello2008} presented in Fig.~\ref{fig:stabilisers} is constructed from a subset of the stabilisers.
        Due to the graph state's symmetry, this results in multiple inequalities that 
        experimentally yield slightly different violations. 
        For the GHZ state $\state{\text{GHZ}_4}$ their values are between  
        \qty{ 3.21 \pm 0.09}{} and \qty{ 3.05 \pm 0.09}{} and for the linear graph state $\state{\text{L}_4}$
        between \qty{ 2.85 \pm 0.09}{} and \qty{ 2.78 \pm 0.10}{}.
        The inequality with the highest violation, shown in Fig.~\ref{fig:stabilisers}, corresponds for state $\state{\text{GHZ}_4}$ to
        \begin{align}
            \exval{\inequal_{\text{II}}} &= \exval{g_1 + g_1g_3 + g_1g_4 + g_1g_3g_4}\\
                                         &= \exval{XXXX} - \exval{YXYX} + \exval{YXXY} +  \exval{XXYY}, \notag
        \end{align} 
         and for $\state{\text{L}_4}$ to
        \begin{align}
            \exval{\inequal_{\text{II}}} &= \exval{g_2 + g_1g_2 + g_2g_3 + g_1g_2g_3}\\
                                         &= -\exval{ZZ11} - \exval{YYZ1} + \exval{XYYZ} -  \exval{YXYZ} \notag.
        \end{align}
        
    \subsection{MBQC operations}
        The generated states are locally equivalent to the graph states required for the presented \gls{mbqc} operations.
        To perform these operations on our four-qubit linear graph state $\state{\text{L}_4}$ and the state $\state{\text{GHZ}_4}$, the necessary local unitary operations were combined with the measurement bases.\\
        Deterministic \gls{mbqc} needs feed-forward of the measurement outcomes $m_i$ to apply outcome-dependent corrections~\cite{raussendorf2001, briegel2009}.
        For the gates and algorithms presented here, deterministic operations are possible by applying these corrections in post-processing, as listed in Tab.~\ref{tab:corrections}.
        \begin{table}[H]
            \centering
            \caption{Measurement-depended corrections for the different MBQC operations.
                     \label{tab:corrections}}
            \begin{tabular}{|l|l|}
            \hline
            MBQC operation & measurement-depended corrections\\
            \hline
            $T$ gate & $X_4^{m_3} Z_4^{m_2}Z_4^{m_1}$\\
            C$Z$ gate & $X_1^{m_2}X_4^{m_3}$ \\
            CNOT gate & $Z_2^{m_4}Z_3^{m_4}$ \\
            Grover's algorithm & $X_2^{m_4} X_3^{m_1}$\\
            Deutsch's algorithm &  $Z_3^{m_4}$\\
            \hline
            \end{tabular}
        \end{table}
        The uncertainties of the \acrfull{mle} reconstructed fidelities and the classification probabilities of the Deutsch-Josza algorithm were calculated using Monte Carlo simulations, with 1000 simulations per data set and assuming Possionian-distributed counts.
        For the Grover's search identification probabilities (see Fig.~\ref{fig:algos}) the uncertainties are based on Gaussian error propagation, again under the assumption that the recorded coincidence-counts exhibit Poissonian-distributed noise.
\end{document}